# Load Transfer along Continuous Collagen Fibers Reduces the Importance of Wall Thickness Variations

**Authors:** Yamnesh Agrawal (1), Masoud Zamani (1), James R. Thunes (2), Spandan Maiti (1,3,4), Anne M. Robertson (1,3)*

**Affiliations:**

(1) Mechanical Engineering & Materials Science, University of Pittsburgh, Pittsburgh, PA, USA

(2) ANSYS Canada Ltd, Waterloo, ON, Canada

(3) Bioengineering, University of Pittsburgh, Pittsburgh, PA, USA

(4) Chemical and Petroleum Engineering, University of Pittsburgh, Pittsburgh, PA, USA

*Corresponding author (rbertson@pitt.edu)

## Abstract:

The mechanical response of biological soft tissues is influenced by wall heterogeneity, including spatial variations in wall thickness. Traditional models for homogeneous soft tissues under uniaxial loading predict higher stretch and stress in thinner regions. In prior studies, the role of collagen fibers in regions of thickness transition has been largely neglected or only considered in terms of their effect on anisotropy. Here, we explore the role of collagen fibers as primary load-bearing components across regions of varying wall thickness, using a three-dimensional meso-scale model (MSM) incorporating explicit collagen fiber architecture and a gradual thickness gradient. We examined two distinct collagen fiber configurations across the thickness transition: one featuring abrupt fiber termination and another with fiber continuity. Finite element analysis (FEA) under uniaxial tension revealed that load transfer by the continuous fibers markedly reduced the importance of the change in wall thickness, with stretch differentials dropping from 20.97% (fiber-termination network) to 0.68% (continuous fibers) and stress differentials dropping from ~65% (fiber-termination network) to 2.3% (continuous fibers). Fiber tortuosity delayed the point at which mechanical response was governed by fiber structure. These findings demonstrate the critical role of fiber continuity in reducing stretch and stress gradients across regions of varying wall thickness and clarify the importance of accurately representing fiber architecture when modeling soft tissues with heterogeneous wall thickness.

## Introduction:

Advances in imaging techniques such as multiphoton microscopy (MPM) have enabled high-resolution visualization of the three-dimensional architecture of collagen fibers in soft tissues. These imaging techniques have revealed the critical importance of microstructural features such as fiber orientation, volume fraction, and tortuosity in determining local mechanical behavior in organs, including arteries, bladder, and cornea [1-7]. However, various pathological conditions such as aneurysms, atherosclerosis and vascular calcifications can disrupt this microstructural organization of the tissue in the form of focal changes wall thickness [8], fiber degradation [9], and abrupt alterations in fiber organization [10, 11]. In particular, regional variations in wall thickness coupled with alterations in fiber organization are frequently observed in both healthy and diseased arterial tissues, yet their impact on the mechanical response remains poorly understood [12-16].

A close examination of the literature reveals two distinct patterns of fiber organization in regions of varying wall thickness, although this distinction has not been explicitly characterized by the experimental mechanics community. In some tissues, such as cerebral aneurysms with clustered microcalcifications, collagen fibers appear to terminate abruptly at the boundaries of microcalcification pockets as shown in Figure 3D of Gade et al. [17] and Figure 4C–D of Gade et al. [18] and noted in the corresponding text. These observations suggest that collagen degradation and local fiber loss may occur in regions of calcification, disrupting fiber continuity at the boundaries of calcification regions and potentially altering local mechanics. In contrast, work from our laboratory demonstrates that collagen fibers can span continuously across thickness gradients such as that shown in Fig. 1b [19]. Despite the visible qualitative difference between these two architectures, no prior study has systematically investigated how the microstructural differences modulate the biomechanical behavior of the tissue.

Furthermore, understanding the mechanical implications of fiber continuity across thickness gradients requires computational models that accurately represent both wall heterogeneity and fiber architecture. Most prior computational models of arterial biomechanics have either assumed uniform wall thickness [20-23] or utilized isotropic material models which cannot capture the anisotropic, fiber dominated response inherent to arterial tissues [24-27]. Using these models and the common idealization of homogeneous material properties, thinner regions will experience elevated deformations and stress under mechanical loading, due to reduced cross-sectional areas, a prediction rooted in classical engineering mechanics principles [28, 29]. While recent studies have used continuum anisotropic formulations to capture collagen fiber behavior with local thickness variations, they did not explicitly represent discrete fibers [30-32]. As a result, they did not account for changes in fiber volume fraction that naturally arise as continuous fibers cross through regions of varying all thickness.

In this study, we aim to address these gaps in knowledge by explicitly comparing two collagen fiber architectures: one incorporating abrupt collagen fiber terminations (such as identified in the pathological aneurysm tissue in Fig. 1a) and the other with continuous collagen fibers that smoothly transition across regions of varying wall thickness (e.g., Fig. 1b). We hypothesize that collagen fiber spatial continuity across varying thickness regions plays a pivotal role in modulating the mechanical response of soft tissues. To test this hypothesis, we employed a 3D meso-scale model incorporating local thickness variations along with discrete collagen fiber architecture. For the finite element analysis (FEA) of both fiber architectures, we leveraged a validated discrete fiber-embedded finite element framework previously developed by our group (Thunes et al. 2016 [33]). Finite element analysis under uniaxial loading was performed on the model to obtain stretch and stress responses. Given prior evidence that fiber tortuosity influences fiber recruitment and load-bearing behavior [5, 33-35], we also examined how fiber waviness modulates mechanical outcomes in regions of thickness transition. Additional parametric analyses were also conducted to evaluate the influence of fiber continuity coupled with other geometric features, including the steepness of the thickness gradient, the orientation angle between dual fiber families, and the ratio of continuous-to-abrupt fibers in mixed fiber network architectures.

Our results consistently demonstrate that fiber continuity plays a critical role in modulating the mechanical behavior of soft collagenous tissues with heterogeneous wall thickness. Models incorporating continuous fibers exhibited more homogeneous mechanical responses compared to those with abruptly terminating fibers. These findings underscore the need for further data sets on collagen fiber continuity in soft tissues, motivating the biomechanics community to incorporate analysis of fiber continuity in future bioimaging studies. These findings also advance our understanding of structure-property function relationships in soft tissues with heterogeneous thickness and provide a better framework for developing more accurate computational models of vascular pathologies.

# 2. Methods

## 2.1 Acquisition and Imaging of arterial microstructure

Examples of the contrasting microstructural organizations that are central to the biomechanical comparisons in this study are represented in the multiphoton images of vascular tissue shown in Fig. 1. An example with abrupt fiber termination is shown in Fig. 1a and a second with continuous fibers spanning the thickness gradient (arrow) is shown in Fig. 1b.

The specimen in Fig. 1a was harvested from a human cerebral aneurysm, imaged using a multiphoton microscopy (MPM) protocol developed by Gade *et al.* 2019 [17]. Collagen fibers terminate at the boundaries of a region filled with clusters of micro-calcifications. The specimen in Fig. 1b was obtained from a cerebral artery in a study of collagen fiber organization around atherosclerotic plaques using MPM imaging, Ramezanpour 2025 [19]. Here, the fibers can be seen to continuously traverse from thick to thin regions around the plaque.

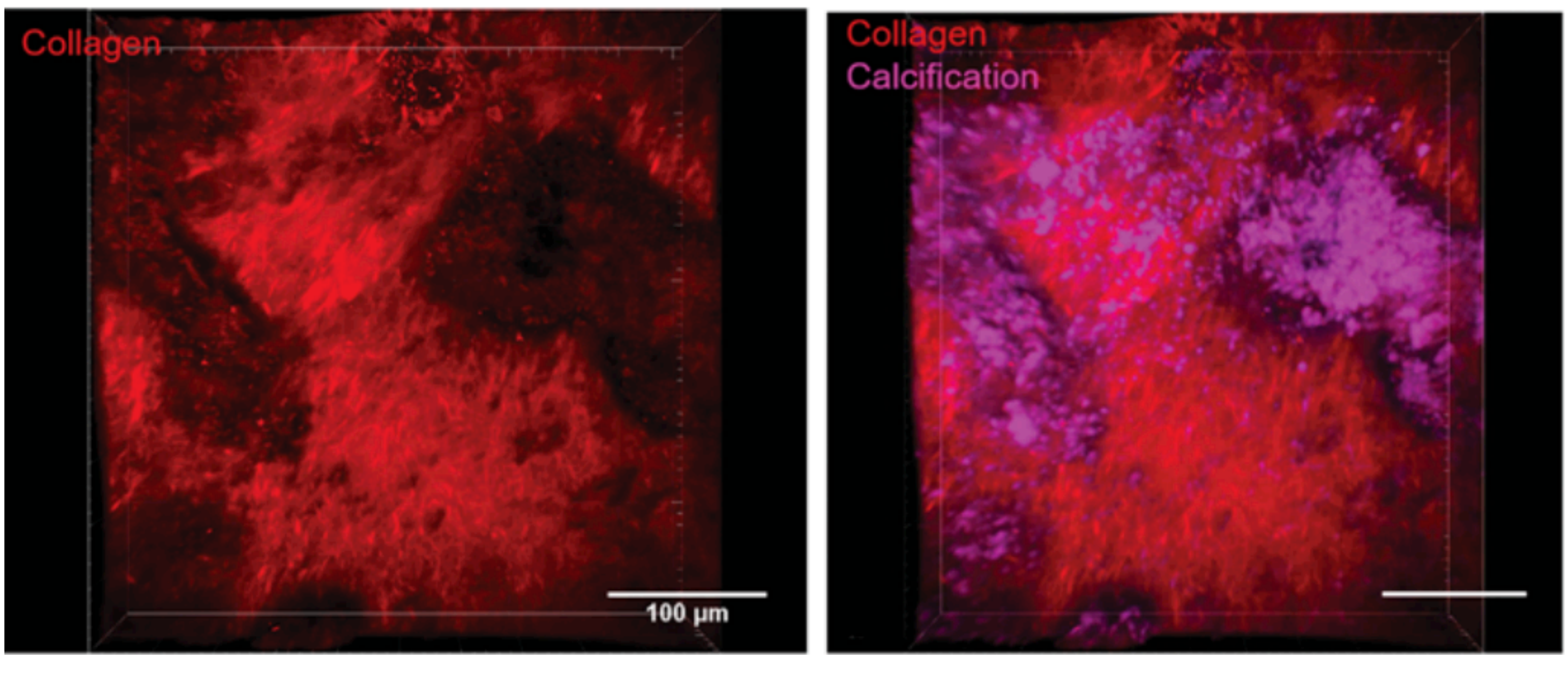

(a)

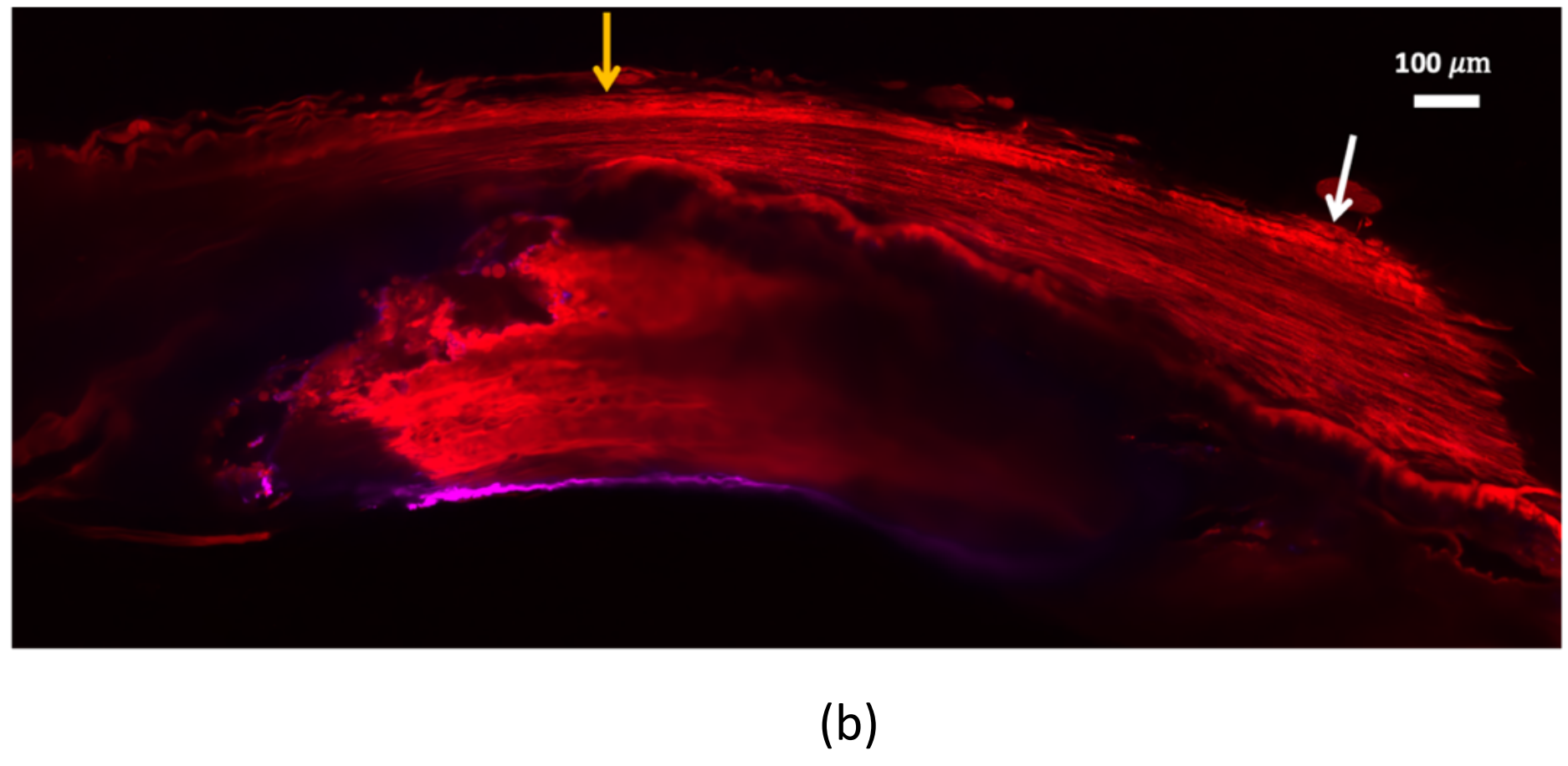

(b)

**Figure 1**: Multiphoton microscopy images showing collagen fiber organization in vascular soft tissues displaying different types of fiber continuity. (a) Enface MPM images from a cerebral aneurysm specimen exhibiting abrupt collagen fiber termination near clusters of microcalcfications. The left panel shows collagen fibers (red) alone, while the right panel includes both collagen fibers (red) and calcification (magenta), highlighting how mineral deposits can disrupt fiber continuity at ECM boundaries (obtained from Gade et al. 2018 [17], with permission). (b) MPM image of a cross section of cerebral artery tissue with atherosclerotic plaque displaying continuous fibers traversing from a thin region (yellow arrow) to thick region (white arrow), (obtained from Ramezanpour, 2025 [19]). All scale bars are 100 $\mu$m. The wavy internal elastic lamina can be seen above the atherosclerotic plaque separating the media from the atherosclerotic plaque.

## 2.2 Model Construction

A three-dimensional meso-scale volume element was developed to investigate the biomechanical effect of the wall thickness variation. The specimen geometry was designed as a rectangular base of dimensions of 500 × 500 µm with thickness varying smoothly from 100 µm (thin region) to 200 µm (thick region) along the uniaxial tensile loading direction (x-axis as shown in Fig. 2a). The thickness profile was mathematically defined using a sigmoid function to ensure gradual transition as shown in Eq. 1.

$$t(x) = t_{\min} + \frac{t_{\max} - t_{\min}}{1 + e^{a(x+b)}} \qquad (1)$$

where $t_{min}$ = 100 µm, $t_{max}$ = 200 µm, and parameters *a* (scaling) and *b* (shifting) controlled the steepness and position of the thickness gradient. For this study we have used a= $5\times10^4$ and b=0 for smooth thickness transition. A planar quadrilateral mesh was first generated in Cubit (Coreform LLC, Orem, UT) and extruded along the thickness direction (z-axis) using a custom Python script to assign spatially varying element thickness based on the sigmoid function. The resulting hexahedral mesh (Fig. 2a) ensured element continuity across the thickness transition region. This idealized geometry was intentionally selected over entire physiological geometries to isolate the influence of fiber continuity from confounding effects introduced by more complex arterial geometries. Such idealized domains have been commonly used in prior studies to enable controlled biomechanical investigations [33, 36, 37]. To evaluate the influence of the thickness transition profile, we investigated two more distinct gradient steepness levels (one smoother and one steeper than current baseline model). The sigmoid slope parameter a (Eq. 1) was specified as $a=2.5\times10^4$ to represent a 'Smoother' transition and as $a=1\times10^5$ to represent a 'Steeper' transition in wall geometry.

## 2.3 Collagen Fiber Network Generation

Collagen fibers were modeled as 1D rod elements where fibers were primarily aligned along the uniaxial loading direction (x-axis as shown in Fig. 2a). To mimic the physiological fiber dispersion in the arterial tissue, the individual fiber orientations were randomly generated from a Gaussian distribution with standard deviation of around 10° [38, 39]. The collagen fiber diameters were set to 1.28 μm based on multiphoton microscopy measurements from a previous study [40]. To understand the effect of collagen fiber continuity in thickness transition regions on the mechanical behavior, two distinct collagen fiber networks were implemented in our meso-scale model. In particular, these network models are:

1. **Abrupt Fiber Termination Network**: In this configuration, collagen fibers were restricted to x-y planes (see Fig. 2a) throughout the MSM thickness and did not continue across the region of thickness variation. A random ray-shooting algorithm was used to generate fibers in which the random seed points were selected within the model domain, and rays were extended from those seed points within the x-y plane with fiber dispersion until they intersected the model boundary. This process was repeated until the desired fiber volume fraction of 35.5% was reached [5]. As a result, fibers originating in the thick region terminated abruptly at the edge of the thickness gradient, producing a discontinuous fiber architecture representative of structurally disrupted tissue (Fig. 2b). Namely, fiber volume fraction was maintained along the x-axis while fiber number diminished as the wall transitioned to a thinner region.

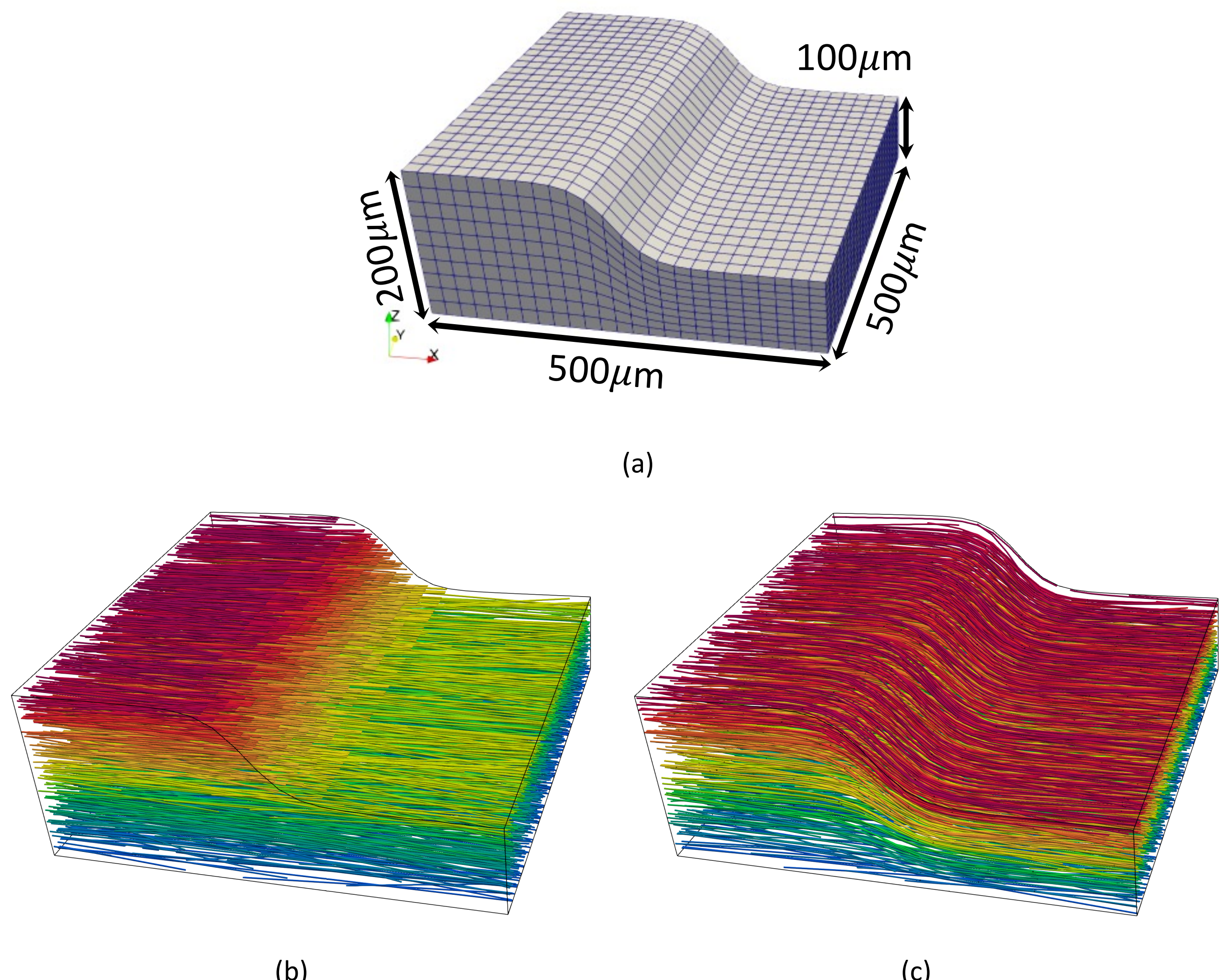

**Figure 2**: Meso-scale model geometry and collagen fiber networks. (a) Hexahedral mesh of the model with a smooth thickness gradient along the x-axis (corresponding to circumferential direction). Thickness transitions from 200 μm (thick region, left) to 100 μm (thin region, right) following a sigmoid function (Eq. 1). The mesh comprises 500×500 μm in-plane dimensions with hexahedral elements. (b) Collagen fiber network shown with abrupt termination at the thickness transition boundary (constant fiber volume fraction). (c) Smoothly transitioning fiber network shown where continuous collagen fibers span the thickness gradient (fiber number is maintained). Fibers adapt spatially to maintain smooth continuity in regions of thickness gradient.

2. **Continuous Fiber Transition Network**: In contrast, this network incorporated fibers that smoothly traversed from the thick to thin region across the thickness gradient, representing fiber continuity across the specimen. During model mesh extrusion along the z-axis, 30 intermediate non-planar surfaces were extracted to capture the smoothly varying thickness profile. Fibers were then generated as connected line segments across these layers, aligned primarily along the x-axis (see Fig. 2a) with angular dispersion to maintain orientation distribution. This method ensured that individual fibers gradually transitioned from thick to thin regions (and vice versa), preserving fiber continuity (and fiber number) across the thickness gradient (Fig. 2c). The total number of fibers was calibrated to achieve a collagen volume fraction of 35.5% in the MSM insuring consistent fiber volume fraction with the abrupt fiber termination network.

## 2.4 Parametric Variations of Fiber Architecture

To evaluate the robustness of our findings beyond these two idealized extremes, additional parametric studies were performed. First, to assess the influence of fiber orientation, we generated networks with dual fiber families symmetric about the loading axis. Three distinct architectures were created with total included angles of 30° (± 15°), 45° (± 22.5°), and 60° (± 30°) relative to the loading direction. This allowed us to determine if the load-transfer mechanism of continuous fibers persists when fibers are not perfectly aligned with the primary load path.

Second, to investigate physiologically realistic intermediate states, we generated "mixed" networks containing both continuous and abruptly terminating fibers. By varying the ratio of continuous fibers from 0% (purely abrupt) to 100% (purely continuous) in increments of 25%, we examined the transition in mechanical behavior. In these mixed models, the total collagen volume fraction was maintained at 35.5%, ensuring that differences in mechanical response were driven solely by the proportion of continuous load paths bridging the thickness gradient.

## 2.5 Finite Element Analysis

The meso-scale model was discretized using a hybrid finite element approach comprising standard 8-noded hexahedral solid elements for the matrix and custom-developed fiber-embedded bar elements to represent collagen segments. This modeling strategy is based on the discrete fiber-embedded finite element framework previously developed and validated by our group (Thunes et al., 2016 [33]). This discrete fiber framework has also been employed in several prior studies to enable fiber-resolved simulations of soft collagenous tissues [23, 36, 41, 42]. In this approach, fiber segments were embedded to reside within the 3D hexahedral matrix element. A key component of this embedding process is the use of kinematically constrained virtual nodes at the ends of each 1D fiber element (as shown in Fig. 3a from Thunes et al 2016 [33]). These nodes are defined in isoparametric coordinates with respect to the parent hexahedral element, thereby ensuring affine deformation of the fiber relative to the surrounding matrix. The constitutive model of the non-fibrous component was defined as the nearly incompressible two-parameter neo-Hookean material model shown in Eq. 2.

$$\Psi = 0.5\mu\left(J^{-\frac{2}{3}}I_1 - 3\right) + 0.5K(J-1)^2 \tag{2}$$

Here, $\Psi$ denotes the strain energy per unit reference volume, $\mu$ represents the shear modulus, J denotes the determinant of the deformation gradient, $K$ is a penalty parameter that enforces the incompressibility condition, and $I_1$ refers to the first invariant of the right Cauchy-Green deformation tensor. The rods were modeled as 1D NeoHookean elements with a recruitment stretch built in, as described in [5, 33]. Briefly, the constitutive model for the fibrous component was defined such that fiber stress ($P_f$) increases linearly with fiber stretch beyond the recruitment stretch $\lambda_r$ with elastic modulus of $E_f$ as shown in Eq. 3 [5]. In this model, a multiplicative decomposition of fiber stretch ($\lambda_f$) has been applied between the true fiber stretch ($\lambda_t$) and recruitment stretch ($\lambda_r$) (i.e., $\lambda_f = \lambda_r\lambda_t$)[37]. In this study, the mechanical response was compared for three recruitment stretches ($\lambda_r$ =1.01, 1.25, and 1.5) in order to understand the combined effect of fiber tortuosity and fiber organization in thickness varying tissues.

$$P_f = \begin{cases} 0 & \text{if } \lambda_f < \lambda_r \Rightarrow \lambda_t < 1 \\ E_f\left(\lambda_f/\lambda_r - 1\right) & \text{if } \lambda_f \geq \lambda_r \Rightarrow \lambda_t \geq 1 \end{cases} \tag{3}$$

To ensure the adequacy of the chosen mesh resolution, a mesh convergence study was performed by approximately doubling the number of elements. The resulting deformation fields showed less than 2% difference between the coarse and refined meshes across both thick and thin regions. These findings confirm that the current mesh density is sufficient for capturing the mechanical response accurately in this study. To simulate uniaxial tensile testing, displacement-controlled boundary were applied on our computational model as per our previous publication [5]. The uniform displacement was applied to the right face (thinner tissue region) to impose axial stretch in the loading direction (x-axis in Fig 2a). To prevent rigid body motion and ensure uniaxial deformation, roller constraints were applied on three orthogonal faces: the left face was restricted in the x-direction, the front face in the y-direction, and the bottom face in the z-direction. Simulations were performed quasi-statically with 2000 load steps to ensure convergence.

## 2.6 Data Analysis

All displacement results were post-processed in Paraview (v5.12, Kitware Inc., Clifton Park, NY) to obtain the localized stretch in the uniaxial loading direction and von Mises's stress fields in all thickness regions. Regional average properties were computed for thick (t = 200 μm) and thin (t = 100 μm) regions of the MSM. Percent differences in average stretches in the loading direction between regions were calculated as:

$$\Delta\lambda(\%) = \frac{\lambda_{thin} - \lambda_{thick}}{\lambda_{thick}} \times 100 \quad (4)$$

# Results:

## 3.1 Stretch and Stress Patterns Under Uniaxial Loading

Finite element simulations of the MSM under uniaxial tensile loading revealed distinct stretch and stress patterns between the abrupt and smooth fiber networks. At an applied tissue stretch of λ=1.5, the abrupt fiber termination network exhibited substantial stretch and stress disparities between thick (t=200 μm) and thin (t=100 μm) regions (Figs. 3a & 4a). The thin region experienced an average stretch of 1.56 ± 0.06, accompanied by a relatively high von Mises stress of 0.56 ± 0.10 MPa. In contrast, the thick region experiences a lower stretch (1.29 ± 0.18) and stress (0.34 ± 0.15 MPa). The maximum von Mises stress across the entire MSM tissue with abrupt fiber network was 2.49 MPa.

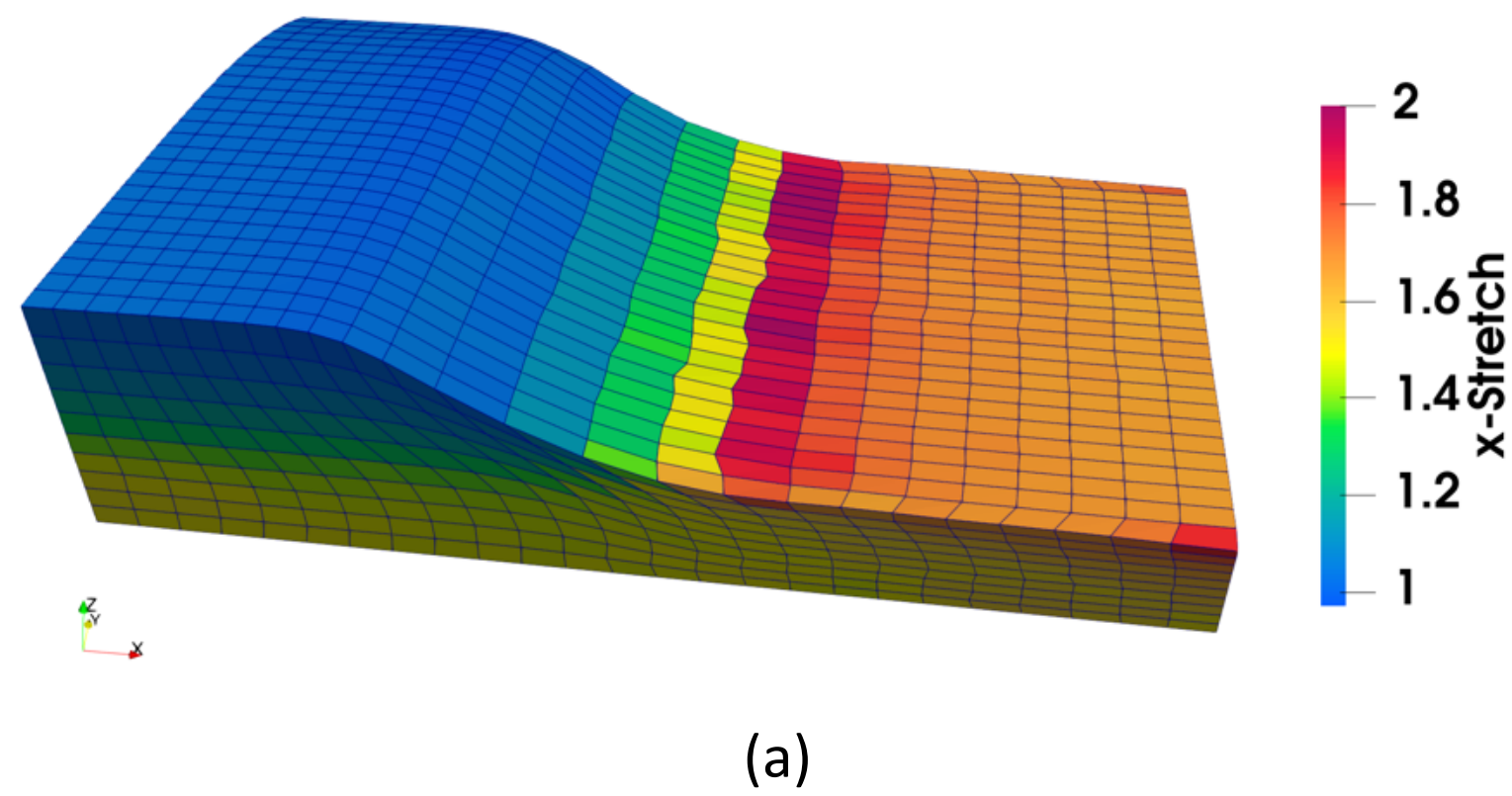

(a)

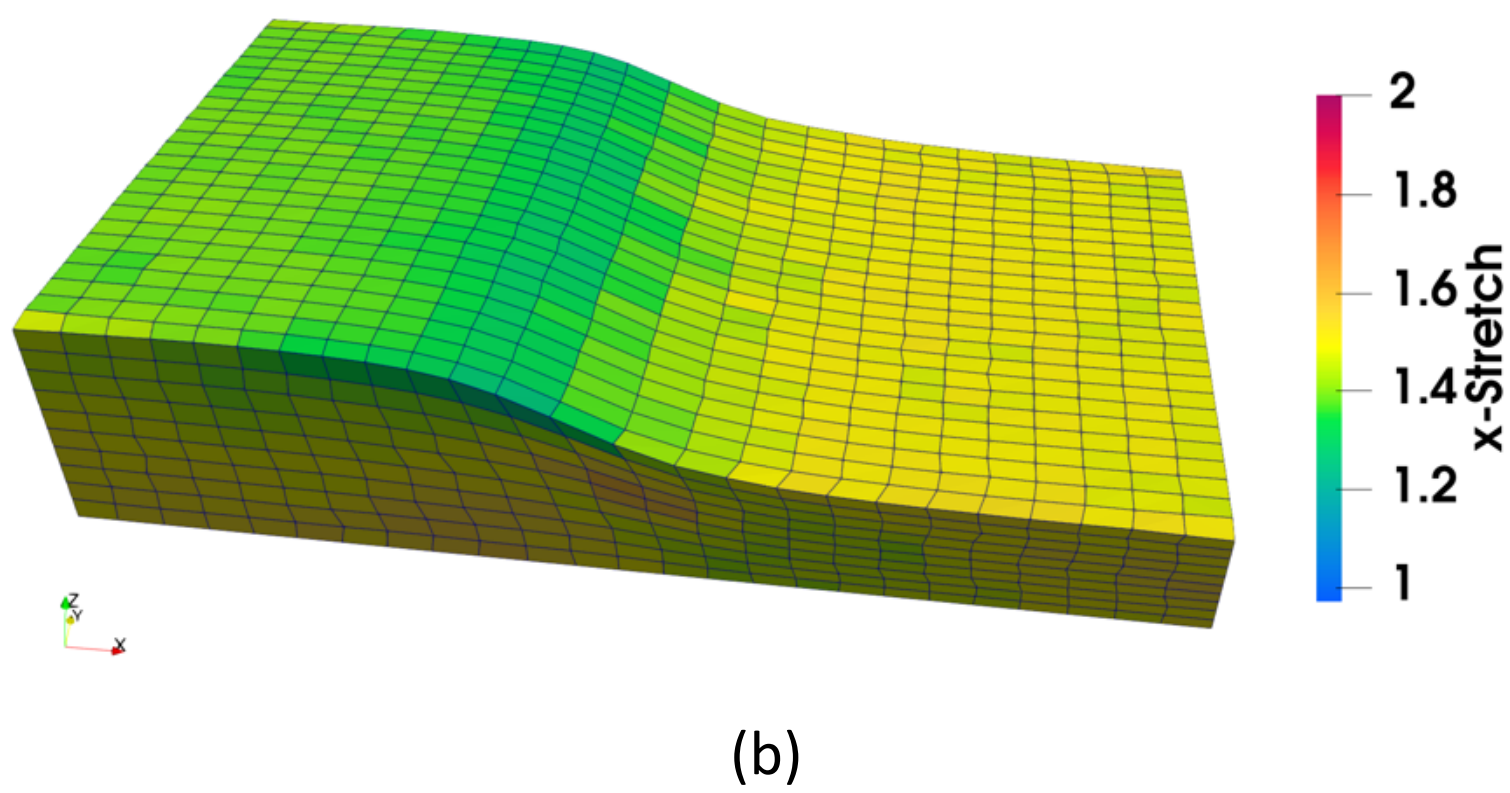

(b)

**Figure 3**: Spatial distribution of tissue stretch in the loading direction at an applied tissue stretch λ = 1.5 for (a) Abrupt fiber network with average stretch of 1.3 ± 0.18 in thick and 1.56 ± 0.06 in thin region. (b) Continuous fiber network with average stretch of 1.47 ± 0.07 in thick and 1.5 ± 0.04 in thin region.

In contrast, the continuously smooth fiber transition network diminished the spatial gradient in the mechanical response (Figs. 3b & 4b). The thin region exhibited an average stretch of 1.48 ± 0.02 and average stress of 0.45 ± 0.02 MPa, while the thick region showed an average stretch of 1.47 ± 0.07 and average stress of 0.44 ± 0.05 MPa. The maximum von Mises stress in the MSM tissue with continuous fiber network was 0.664 MPa. The percent difference in stretch between both thickness regions decreased from 20.97% for the abrupt fiber termination network to 0.68% for the continuous fiber network.

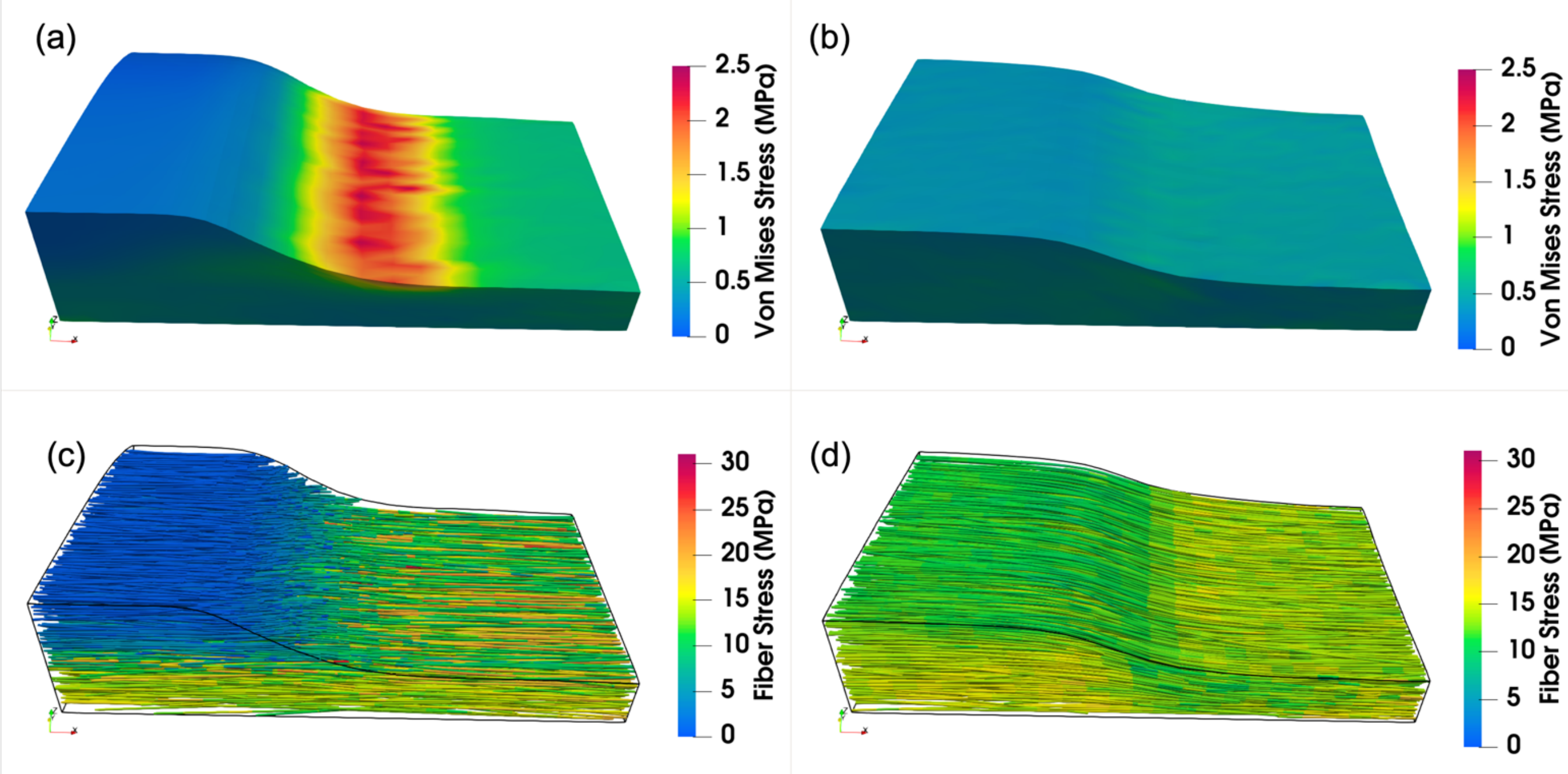

**Figure 4:** Effect of collagen fiber continuity on matrix and fiber stress distributions in soft tissues with wall thickness variation. Spatial distribution of matrix von Mises stress at applied tissue stretch $\lambda = 1.5$ for (a) Abrupt fiber network with average stress of 0.35 ± 0.15 MPa in thick and 0.57 ± 0.18 MPa in thin regions. (b) Continuous fiber network with average stress of 0.44 ± 0.05 MPa in thick and 0.45 ± 0.16 MPa in thin regions. (c, d) Fiber-level stress distribution for the same two architectures. In the abrupt network (c), the maximum fiber stress reaches 30 MPa, with an uneven stress distribution characterized by lower stress in the thick region. In contrast, the continuous network (d) exhibits a lower maximum fiber stress of 20 MPa. Fibers in this network transition smoothly between thick and thin regions, facilitating consistent load transfer.

Fiber-level stresses at $\lambda = 1.5$ are shown in Fig. 4c,d. In the abrupt network, collagen fibers in the thin region experienced substantially higher stress (14.32 ± 2.08 MPa) compared to those in the thick region (7.67 ± 6.15 MPa) (Fig. 4c). The continuous network exhibited a more uniform fiber stress distribution, with the thin and thick regions experiencing similar values of 9.83 ± 6.23 MPa and 9.55 ± 6.09 MPa, respectively (Fig. 4d).

## 3.2 Parametric Analysis of Different Collagen Fiber Architectures

To evaluate the sensitivity of the mechanical response to model parameters, we performed three parametric analyses detailed in Supplementary Notes S1–S3. The results of these analyses regarding the stretch differential between thick and thin regions are summarized in Figure 5.

First, we varied the thickness gradient steepness and obtained their stretch differential (Fig. 5a). In the abrupt fiber termination networks, increasing the gradient steepness from Gradual to Steep resulted in a consistently high stretch differential, hovering around 21%. In contrast, for the continuous fiber networks, the stretch differential remained consistently low (< 1%) regardless of whether the thickness transition was gradual or steep, increasing only marginally from 0.53% to 0.79% with steepness.

Second, we varied the fiber orientation using dual-family networks. In addition to the 0° mean orientation used in the primary analysis, we simulated networks with included angles of 30°, 45°, and 60° (Fig. 5b). As the fiber angle relative to the loading axis increased (reducing axial alignment), the stretch differential in the abrupt networks decreased slightly (from ~21% to ~18%) but remained significantly higher than in the corresponding continuous networks. In the continuous fiber networks, the stretch differential remained below 1% for all angles, with only a marginal increase observed towards higher orientation.

Third, we simulated mixed fiber networks with varying ratios of continuous-to-abrupt fibers (Fig. 5c). The stretch differential exhibited a monotonic, approximately linear decrease as the percentage of continuous fibers increased. The stretch disparity dropped from 20.97% in the purely abrupt network (0% continuous) to 10.97% in the 50% mixed network, and finally to 0.68% in the purely continuous network (100%).

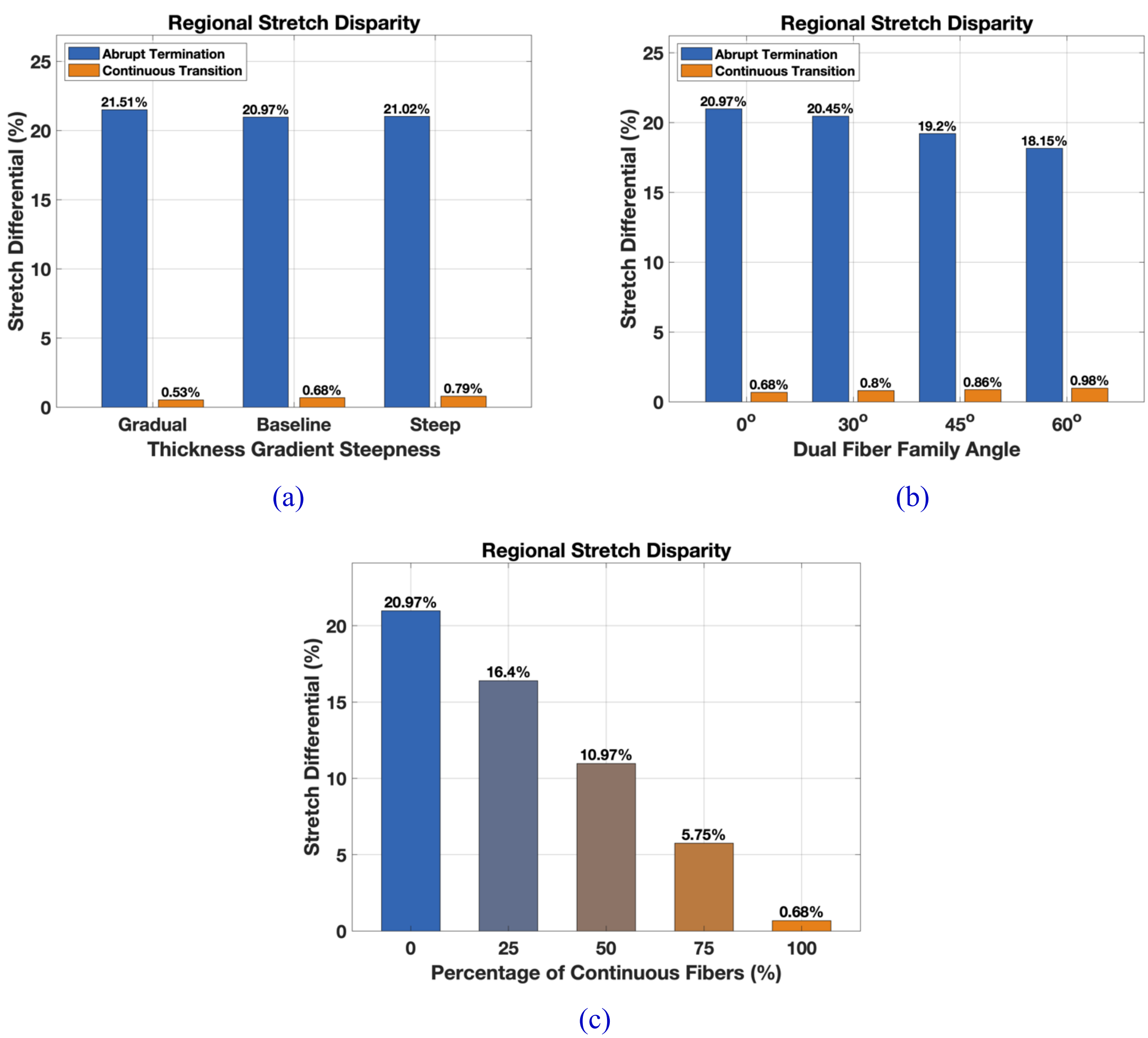

**Figure 5**: Parametric analysis of the fiber continuity effect across varying model parameters. Bar plots show the percentage difference in average stretch between thick and thin regions (obtained from Eq. 4) under uniaxial loading ($\lambda = 1.5$). (a) Influence of thickness gradient steepness: gradual ($a = 2.5\times10^4$), baseline ($a = 5\times10^4$), and steep ($a = 1\times10^5$). (b) Influence of fiber orientation: Single fiber family (0°) and dual fiber families with total angle between them set to 30°, 45°, and 60°. (c) Influence of varying the ratios of continuous-to-abrupt fibers (0% to 100%). In all cases, the presence of continuous fibers (orange bars in a, b; increasing percentage in c) consistently reduces the geometric stretch disparity compared to abrupt fiber terminations, demonstrating that even partial fiber continuity confers significant mechanical benefits.

### 3.3 Local Collagen Fiber Volume Fraction

For the continuous fiber case, the fiber number is constant, so the fiber volume fraction will naturally vary across the thick and thin regions in specimens, Fig. 1b. In particular, while the average collagen volume fraction for the MSM was maintained at 35.5% for both fiber architectures, local volume fraction variations emerged due to differences in fiber arrangement. To quantify this, a local fiber volume fraction analysis was performed, Fig. 6. In the abrupt fiber termination model, the fiber volume fraction was 35.31 ± 5.02% in the thin region and 35.20 ± 3.61% in the thick region (Fig. 6a). In contrast, in the continuous fiber network model, the same number of fibers traversed the MSM, resulting in a higher fiber volume fraction in the thin region (51.93 ± 2.19%) compared with the thick region (26.04 ± 1.10%), Fig. 6b.

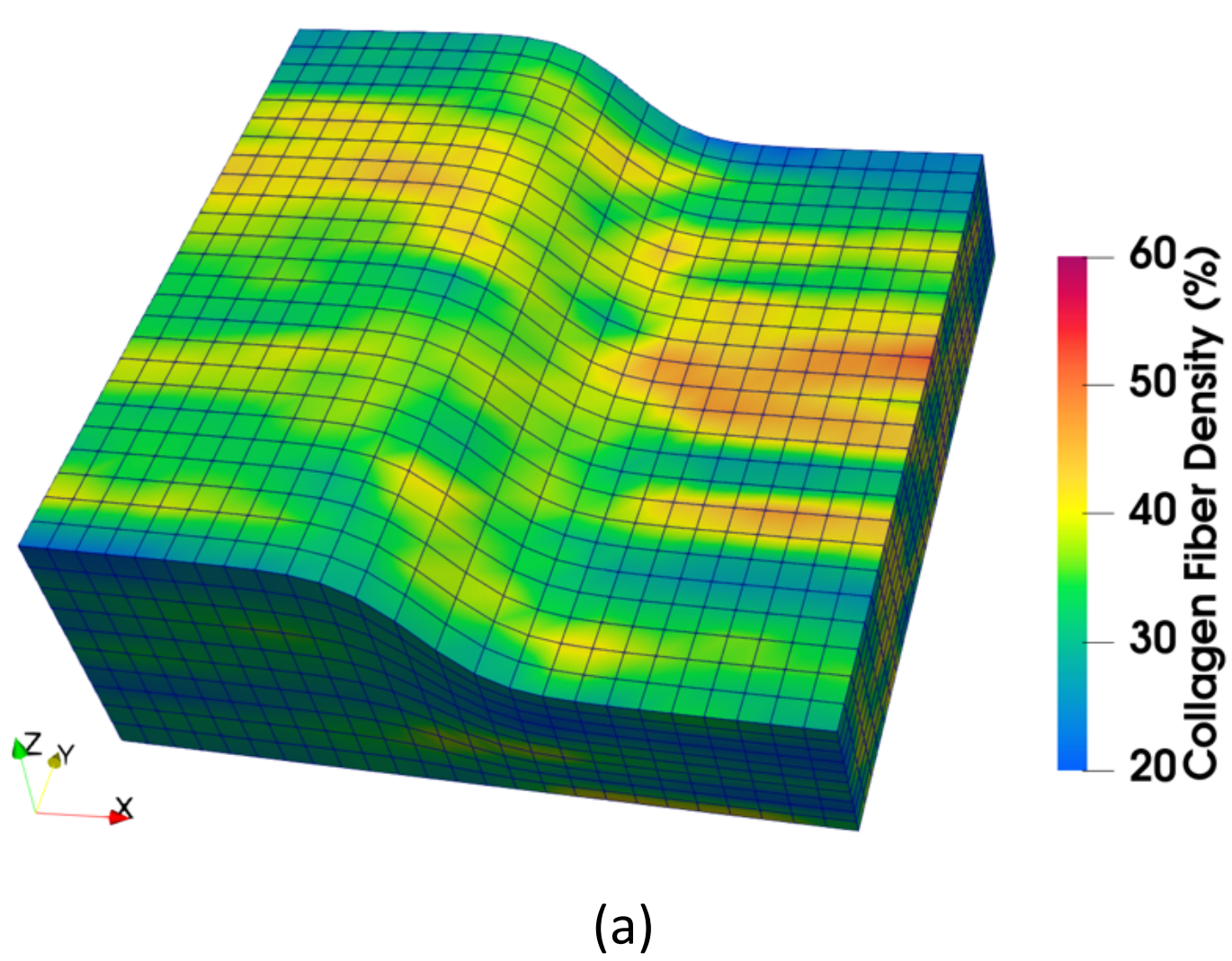

(a)

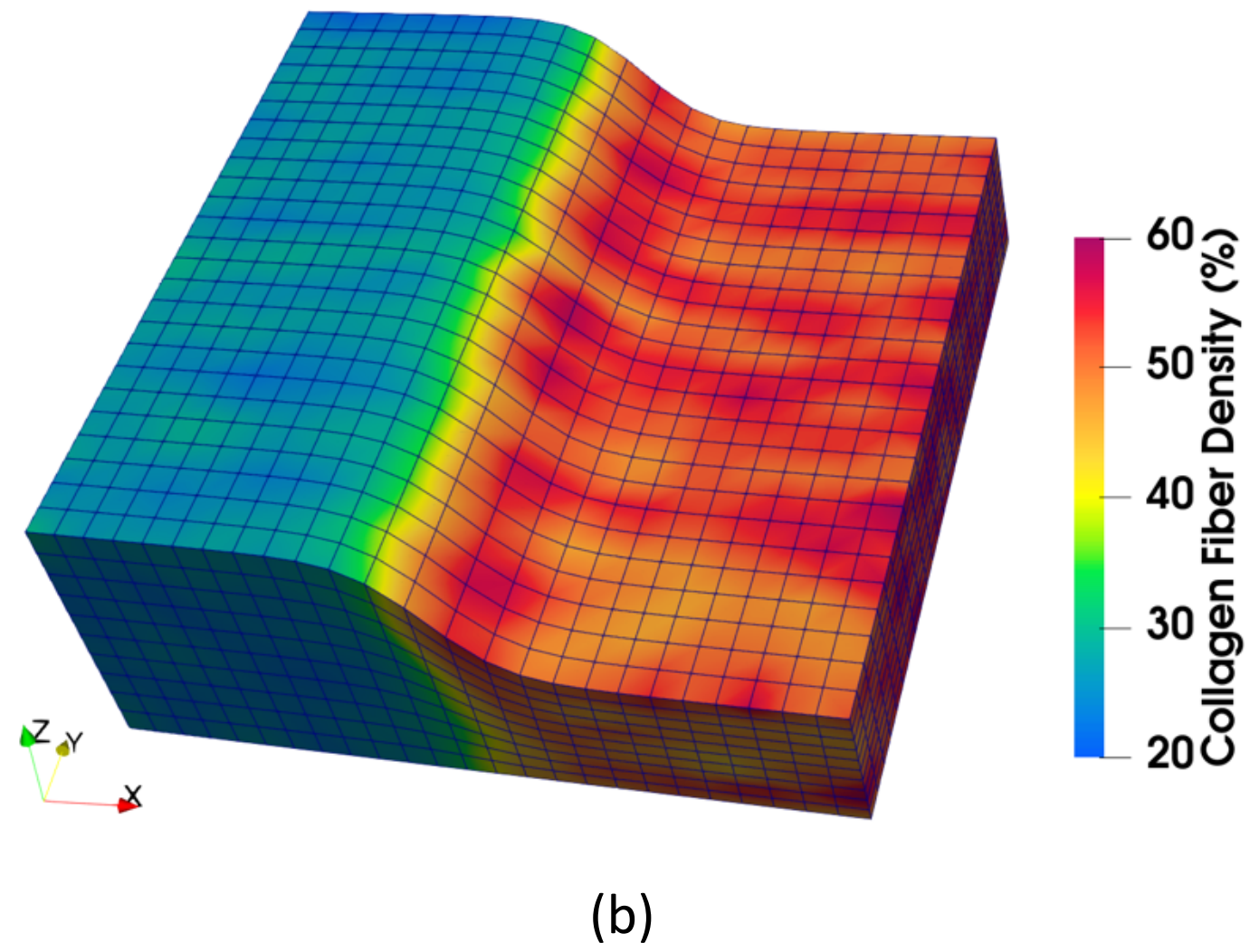

(b)

**Figure 6:** Distribution of collagen fiber volume fraction in our meso-scale model for (a) Abrupt fiber termination network and (b) Continuous fiber transition network. In the abrupt network (a), a fiber volume fraction of 35.5% was prescribed. In contrast, as the continuous network (b) preserves the same total number of fibers across all thickness regions, the thinner region has higher fiber volume fraction due to its smaller wall thickness.

## 3.4 Influence of Fiber Tortuosity on Differences in Deformation and Stress Fields

Fiber tortuosity (controlled through specification of recruitment stretch, $\lambda_r$) substantially modulated the low stretch deformation behavior (Fig. 7). For $\lambda_r$ = 1.01, fibers start engaging almost immediately after applying the external load. Consequently, for the abrupt fiber termination network, the stretch difference in thick and thin regions keeps widening, reaching 32.5% at applied tissue stretch of two. For the continuous fiber network, there is a marginal stretch difference of only 2.27% at the applied tissue stretch of two.

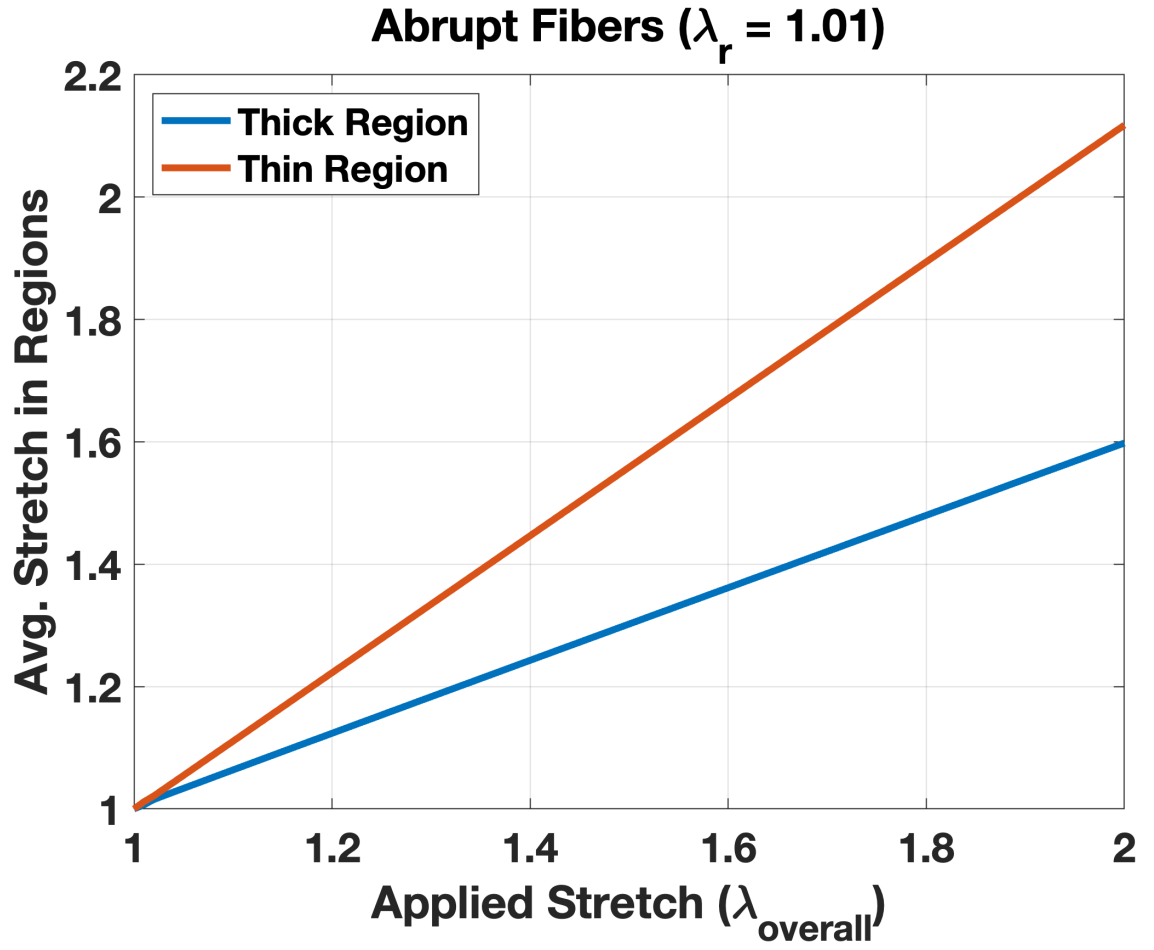

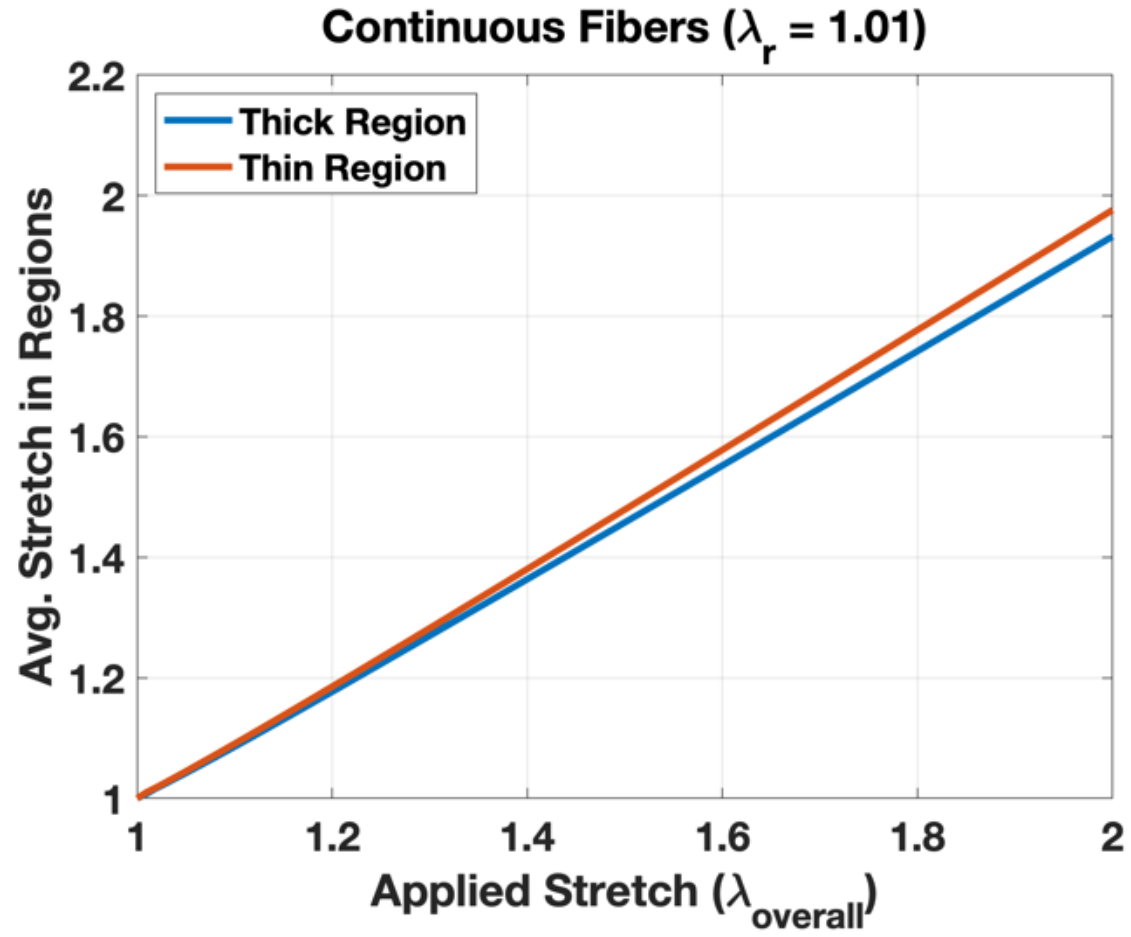

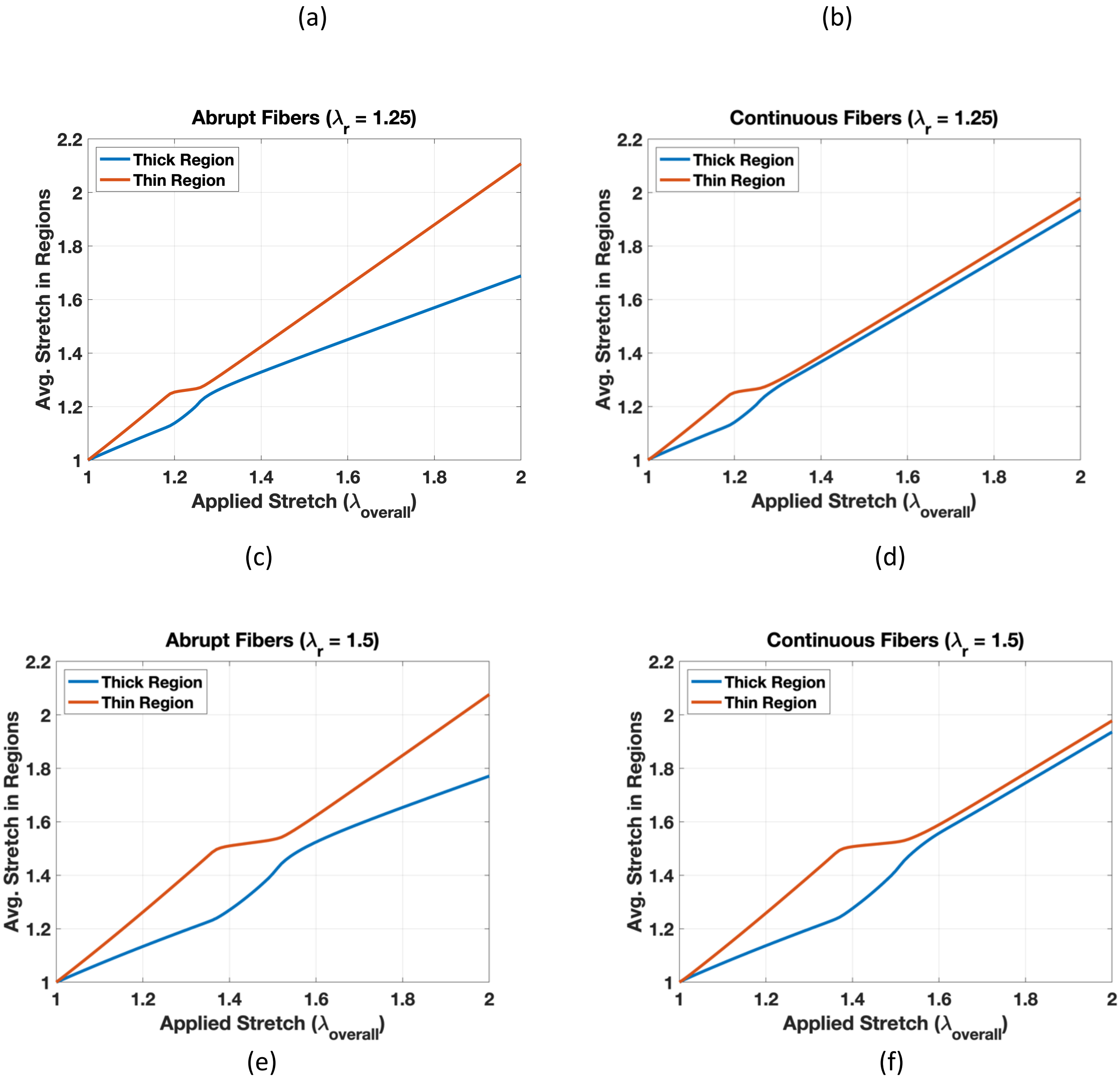

**Figure 7**: Deformation disparities between thick and thin regions as a function of applied tissue stretch. Average stretch in thick and thin regions for abrupt and smooth fiber networks at tortuosity thresholds $\lambda_r = 1.01$ (almost no tortuosity) (a-b), $\lambda_r = 1.25$ (c-d), and $\lambda_r = 1.5$ (e-f).

For $\lambda_r = 1.25$, fibers engaged later due to higher undulation associated with the larger recruitment stretch. Prior to fiber recruitment in both thick and thin regions, ($\lambda < 1.25$), the abrupt fiber termination network showed a 10.56% higher stretch in the thin region compared to the thick region. In contrast, once fibers in both regions were activated ($\lambda \geq 1.25$), the stretch difference reduced to 3.75% as primary load transfer shifted from the underlying matrix to the stiffer collagen fibers. Similarly, in the continuous fiber network, the disparity decreased from 10.06% (pre-recruitment) to 1.49% post-recruitment.

Moreover, for the abrupt fiber termination network, post-recruitment differences in average stretch between thick and thin regions began to re-emerge at higher values of applied tissue stretch. For example, the stretch disparity increased to 24.82% at $\lambda = 2$. In contrast, the continuous fiber network maintained a smaller difference (2.3%) between thick and thin regions.

At even higher tortuosity ($\lambda_r = 1.5$), fiber recruitment was further delayed. Prior to fiber engagement, the abrupt network showed a 20.77% stretch difference, which decreased to 5.98% after recruitment. For the continuous network, the disparity similarly dropped from 20.16% pre-recruitment to 1.94% post-recruitment. Notably, in the abrupt network, stretch differences re-emerged at higher loading levels post-recruitment, increasing again to 17.24% at $\lambda = 2$. In contrast, the continuous network maintained a minimal stretch differential of 2.17%, highlighting its mechanical robustness even at higher deformations.

These trends are further illustrated in the stress–stretch (SS) curves (Supplemental Figure S4.1), which show that higher $\lambda_r$ values result in flat initial regions of the curve—indicating minimal fiber contribution prior to recruitment. Consequently, tissues with higher fiber tortuosity exhibited lower overall stiffness in early deformation stages, as load was primarily borne by the softer matrix. After recruitment, the fibers significantly influenced the mechanical response.

# 4. Discussion

## 4.1 Collagen Fiber Continuity Mitigates Geometry-Induced Stretch and Stress Disparities

The finite element simulations demonstrated that when collagen fibers smoothly transition through regions with a thickness gradient, they substantially attenuate the stretch disparity between thin and thick regions of the arterial wall. In networks with abrupt fiber terminations, thin regions experienced markedly elevated stretches, reaching up to 21% higher strain compared to thicker counterparts at the overall tissue stretch of two. This result confirms intuitive mechanical expectations for homogeneous materials, wherein thinner regions have greater stress and strain —due to diminished area and larger stress. However, when collagen fibers were allowed to transition smoothly across thickness gradients, the disparity dropped dramatically to 0.68% under the same loading conditions. This more homogeneous strain distribution can be attributed to the ability of the continuous fibers to transfer load from thick to thin regions, thereby redistributing mechanical stresses across the tissue. While the examples shown in Fig. 1 are for vascular tissues, these findings are applicable more generally to any collagenous soft tissues.

Furthermore, different parametric analysis regarding the steepness of the thickness gradient, fiber orientation (dual fiber families), and the proportion of continuous fibers in mixed networks, mentioned in Section 3.2 strengthen the effect of fiber continuity behavior. These studies confirmed that the homogenizing effect of fiber continuity is robust and not limited to the specific parameters chosen for the primary analysis. Whether the thickness transition was gradual or steep, whether fibers were oriented in dual families with angles varying from 30° to 60°, or whether the network consisted of a mixture of continuous and abrupt fibers, the presence of continuous load paths consistently mitigated the stretch and stress disparities. This indicates that the load transfer mechanism provided by fiber continuity is a fundamental mechanical characteristic that overrides geometric stress concentrations across a broad design space.

Interestingly, Cavinato *et al.* (2019) have shown that there is no significant correlation between thickness variations and ascending thoracic aneurysm rupture [43]. Such findings could be explained if collagen fibers smoothly transitioned across the thickness gradients, suggesting a physiological role for fiber continuity in mechanical resilience. Notably, the modeling efforts or clinical interpretations based on geometric parameters alone may be inappropriate for estimating local mechanical vulnerabilities, especially in pathological settings like aneurysms or dissections.

These findings provide compelling motivation for the experimental community to focus on fiber continuity in future studies. While disparate features of the collagen fiber network such as fiber orientation, dispersion, and tortuosity have been extensively quantified in the literature, this specific structural feature of fiber continuity across thickness gradients has been overlooked. Our work identifies this structural feature as a distinct, biomechanically meaningful parameter that significantly influences stress and stretch distribution of the arterial wall. Results from biomechanical studies addressing the role of fiber continuity should provide important mechanistic insights, for example, for understanding tissue failure mechanisms in pathological states.

## 4.2 Distribution of Fiber Volume Fraction and Implications for Load Sharing

Variations in fiber volume fraction arise naturally when fibers smoothly transition between regions of different thickness, as illustrated in Fig. 6. Therefore, fiber volume fraction is directly coupled to fiber organization (abruptly terminated versus continuous) plays a central role in shaping local tissue mechanics. In the abrupt fiber network (Fig. 6a), fibers terminate at the transition zone between thick and thin regions, resulting in more fibers occupying the thicker region and fewer in the thinner one. The reduced number of fibers in the thin region diminishes its load bearing contributions, amplifying local stress and stretch.

By contrast, in the continuous fiber network, the same number of fibers span all thickness regions due to fiber continuity. As wall thickness decreases, these fibers are distributed through a smaller cross-sectional area, increasing local fiber volume fraction in the thin region (Fig. 6b). This densification enhances the mechanical support of the thin region, facilitating improved load sharing capability and contributing to a more uniform deformation response across the tissue.

This behavior can be further understood using a simplified mechanical model. Once fibers are recruited, the effective total fiber stiffness $E$ becomes proportional to the number of engaged fibers N. In continuous networks, N remains constant across thick and thin regions, leading to geometry-insensitive mechanical behavior after recruitment, even in tissues with thickness gradients. However, in abruptly terminated fibers cases, number of fibers in different regions proportionally increase with the local thickness which makes the biomechanical behavior high dependent on the geometry.

It is also important to note the continuum anisotropic models in which fiber volume fraction is held constant across a thickness gradient, have inherently assumed fiber volume fraction is constant, which would correspond to truncated fibers in the underlying matrix.  It should be noted the continuous fiber architecture can still be represented within a continuum anisotropic framework without explicitly modeling discrete fibers. This can be achieved by modulating the local fiber volume fraction as inversely proportional to the local wall thickness, corresponding to preservation of total fiber count across the thickness gradient.

### 4.3 Fiber Tortuosity Introduces Transitional Mechanical Behavior

The simulations further revealed that fiber tortuosity also substantially modulates the local stretch distribution. In the pre-recruitment phase ($\lambda < \lambda_r$), the mechanical response was predominantly governed by the geometry of the MSM as the isotropic matrix was the sole load bearing component. As a result, there were substantial stretch disparities in regions of differing wall thickness for both fiber architectures. For instance, in the continuous fiber network with $\lambda_r = 1.25$, the thin region stretched 10.06% more than the thick region prior to fiber engagement. Once the fiber stretch exceeded the recruitment threshold in both regions, collagen fibers began to engage and assume the primary load-bearing role, substantially reducing the stretch disparity to below 5%. This highlights the dual influence of fiber tortuosity: initially delaying fiber recruitment and amplifying geometry-driven deformation, and later facilitating redistribution of stretch and stress as fibers straighten and stiffen.

Following fiber recruitment, the mechanical response becomes predominantly governed by fibers. However, even in the continuous fiber network, minor stretch disparities in both regions persist at higher stretches, though these are substantially smaller compared to the abrupt fiber termination network. This difference indicates that, while continuous fiber network enhances mechanical uniformity, geometric effects are not entirely eliminated. This can be attributed to loading within the isotropic matrix in which thicker regions have lower stress due to the larger cross-sectional area.

Importantly, since failure in arterial tissues is generally initiated by fiber rupture—which typically occurs only after fibers are recruited—a constant-thickness idealization may be reasonable for failure studies if fiber continuity can be confirmed, even when detailed local thickness maps are unavailable. In such scenarios, once the fibers engage, the geometry-driven stress disparities diminish, and the dominant fiber-governed mechanics override the effects of wall thinning. Consequently, a simplified constant-thickness model should capture the relevant mechanics with reasonable accuracy, provided the underlying fiber architecture is continuous. This conjecture can be evaluated in future biomechanical studies of soft collagenous tissues.

### 4.4 Limitations and Future Directions

While this study provides valuable mechanistic insight, there are some limitations in the study that drive future directions of investigation. First, an idealized geometry was chosen with smooth, mathematically defined thickness gradients and simplified boundary conditions. This idealization was chosen to allow a controlled investigation of the collagen fiber organization. In future studies, we will build on the present work to investigate the additional effects of geometric complexity and heterogeneity found in tissues such as cerebral aneurysms and atherosclerotic arteries considering entire geometry. Future studies will also explore the role of residual stress states and multiaxial in vivo loading conditions.

Moreover, although the collagen fiber network was generated using physiological parameters, experimental validation of the simulated mechanical response, including regional stretch and stress distributions, is an important area for future investigations. The effect of fiber continuity could be validated using mechanical testing with local strain measurement techniques such as digital image correlation (DIC), then compared with FEM results to further strengthen the hypothesis.

The MPM images from previous studies have revealed both types of fiber organization [17-19]. In some cases, abrupt fiber termination occurred near micro-calcified regions (Fig. 1a), while in others, fibers appeared continuous even in diseased tissues with features such as atherosclerotic plaques (Fig. 1b). This suggests that vascular pathologies can also have continuous collagen architectures, and that the presence of disease does not necessarily correspond to complete disruption of fiber networks. This study provides motivation to the biomechanics community to further explore questions of fiber continuity in collagenous soft tissues in both health and disease.

## 5. Conclusion

This study highlights the critical role of continuity of fibers in modulating the mechanical behavior of collagenous soft tissues with wall thickness variations. Finite element simulations revealed that stretch and stress disparities between thick and thin regions can be ameliorated in specimens with continuous fibers spanning regions of thickness gradients, through effective load transfer and localized increases in fiber volume fraction. Fiber tortuosity modulates this behavior by delaying the influence of collagen fibers to higher stretches. As a result, before fiber recruitment, the load is borne by the underlying matrix and hence dominated by the effect of local wall thickness. These findings demonstrate that geometry alone cannot explain local mechanics. Rather, microstructural fiber organization plays a dominant role once fibers are recruited. Therefore, accurate representation of collagen structure is vital for reliable biomechanical modeling and rupture risk prediction, underscoring the need for imaging tools that capture microstructural features such as fiber continuity and local fiber volume fraction. We conjecture that for tissues with continuous fibers, constant-thickness models may yield accurate predictions, even when local thickness data are unavailable.

## Acknowledgements

Support from NIH grant **2R01NS097457** is gratefully acknowledged. The authors also thank Dr. Shaghayegh Zamani Ashtiani for her insightful discussions on computational implementation that contributed to the work. Finally, the authors sincerely thank the anonymous reviewers for their thorough evaluation and constructive comments.

## REFERENCES

1. Tsamis, A., J.T. Krawiec, and D.A. Vorp, *Elastin and collagen fibre microstructure of the human aorta in ageing and disease: a review.* Journal of the royal society interface, 2013. **10**(83): p. 20121004.
2. Cheng, F., et al., *Layer-dependent role of collagen recruitment during loading of the rat bladder wall.* Biomechanics and modeling in mechanobiology, 2018. **17**: p. 403-417.
3. Ji, F., et al., *A direct fiber approach to model sclera collagen architecture and biomechanics.* Experimental eye research, 2023. **232**: p. 109510.
4. Islam, M.R., et al., *Fibrous finite element modeling of the optic nerve head region.* Acta biomaterialia, 2024. **175**: p. 123-137.
5. Agrawal, Y., et al., *Effect of Collagen Fiber Tortuosity Distribution on the Mechanical Response of Arterial Tissues.* Journal of Biomechanical Engineering, 2025. **147**(2).
6. Tan, Y.Q., et al., *Multiphoton microscopic study of the renal cell carcinoma pseudocapsule: implications for tumour enucleation.* Urology, 2020. **144**: p. 249-254.
7. Azari, F., et al., *Elucidating the high compliance mechanism by which the urinary bladder fills under low pressures.* Scientific Reports, 2025. **15**(1): p. 25356.
8. Alagan, A.K., et al., *Histopathology-based near-realistic arterial wall reconstruction of a patient-specific cerebral aneurysm for fluid-structure interaction studies.* Computers in Biology and Medicine, 2025. **185**: p. 109579.
9. Petsophonsakul, P., et al., *Role of vascular smooth muscle cell phenotypic switching and calcification in aortic aneurysm formation: Involvement of vitamin K-dependent processes.* Arteriosclerosis, thrombosis, and vascular biology, 2019. **39**(7): p. 1351-1368.
10. Fortunato, R., et al., *Effect of macro-calcification on the failure mechanics of intracranial aneurysmal wall tissue.* Experimental mechanics, 2021. **61**: p. 5-18.
11. Blaser, M.C. and E. Aikawa, *Roles and regulation of extracellular vesicles in cardiovascular mineral metabolism.* Frontiers in cardiovascular medicine, 2018. **5**: p. 187.
12. Harteveld, A.A., et al., *Ex vivo vessel wall thickness measurements of the human circle of Willis using 7T MRI.* Atherosclerosis, 2018. **273**: p. 106-114.
13. Burke, G.L., et al., *Arterial wall thickness is associated with prevalent cardiovascular disease in middle-aged adults: the Atherosclerosis Risk in Communities (ARIC) Study.* Stroke, 1995. **26**(3): p. 386-391.
14. Acosta, J.M., et al., *Effect of aneurysm and patient characteristics on intracranial aneurysm wall thickness.* Frontiers in Cardiovascular Medicine, 2021. **8**: p. 775307.
15. Dinenno, F.A., et al., *Age-associated arterial wall thickening is related to elevations in sympathetic activity in healthy humans.* American Journal of Physiology-Heart and Circulatory Physiology, 2000. **278**(4): p. H1205-H1210.
16. Duggirala, R., et al., *Genetic basis of variation in carotid artery wall thickness.* Stroke, 1996. **27**(5): p. 833-837.
17. Gade, P.S., A.M. Robertson, and C.Y. Chuang, *Multiphoton imaging of collagen, elastin, and calcification in intact soft-tissue samples.* Current protocols in cytometry, 2019. **87**(1): p. e51.
18. Gade, P.S., et al., *Calcification in human intracranial aneurysms is highly prevalent and displays both atherosclerotic and nonatherosclerotic types.* Arteriosclerosis, thrombosis, and vascular biology, 2019. **39**(10): p. 2157-2167.

19. Ramezanpour, M., *Unraveling the biomechanical impact of collagen and calcifications in vascular tissues using deep learning, high-resolution bioimaging, and advanced computational simulations*. 2025, University of Pittsburgh.
20. Toungara, M., et al., *Micromechanical modelling of the arterial wall: influence of mechanical heterogeneities on the wall stress distribution and the peak wall stress.* Computer Methods in Biomechanics and Biomedical Engineering, 2013. **16**(sup1): p. 22-24.
21. Tricerri, P., et al., *Fluid-structure interaction simulations of cerebral arteries modeled by isotropic and anisotropic constitutive laws.* Computational Mechanics, 2015. **55**: p. 479-498.
22. Holzapfel, G.A. and R.W. Ogden, *Modelling the layer-specific three-dimensional residual stresses in arteries, with an application to the human aorta.* Journal of the Royal Society Interface, 2010. **7**(46): p. 787-799.
23. Mahutga, R.R., et al., *A multiscale discrete fiber model of failure in heterogeneous tissues: Applications to remodeled cerebral aneurysms.* Journal of Biomechanics, 2025. **178**: p. 112343.
24. Scotti, C.M., et al., *Fluid-structure interaction in abdominal aortic aneurysms: effects of asymmetry and wall thickness.* Biomedical engineering online, 2005. **4**: p. 1-22.
25. Shamloo, A., M.A. Nejad, and M. Saeedi, *Fluid–structure interaction simulation of a cerebral aneurysm: Effects of endovascular coiling treatment and aneurysm wall thickening.* Journal of the mechanical behavior of biomedical materials, 2017. **74**: p. 72-83.
26. Torii, R., et al., *Influence of wall thickness on fluid–structure interaction computations of cerebral aneurysms.* International Journal for Numerical Methods in Biomedical Engineering, 2010. **26**(3-4): p. 336-347.
27. Valeti, C., et al., *Influence of wall thickness on the rupture risk of a patient-specific cerebral aneurysm: A fluid–structure interaction study.* Physics of Fluids, 2024. **36**(9).
28. Shang, E.K., et al., *Wall Thickness Influence on Computational Wall Stress of Arteries*. 2012, Lippincott Williams & Wilkins.
29. Muyupa, E., *Measurement of deformation in varying stress fields*. 2018: Open University (United Kingdom).
30. Ma, B., et al., *Nonlinear anisotropic stress analysis of anatomically realistic cerebral aneurysms.* Journal of biomechanical engineering, 2007. **129**(1): p. 88-96.
31. Teixeira, F.S., et al., *Modeling intracranial aneurysm stability and growth: an integrative mechanobiological framework for clinical cases.* Biomechanics and modeling in mechanobiology, 2020. **19**: p. 2413-2431.
32. Sun, Y., et al., *Modeling Fibrous Tissue in Vascular Fluid–Structure Interaction: A Morphology-Based Pipeline and Biomechanical Significance.* International Journal for Numerical Methods in Biomedical Engineering, 2025. **41**(1): p. e3892.
33. Thunes, J.R., et al., *A structural finite element model for lamellar unit of aortic media indicates heterogeneous stress field after collagen recruitment.* Journal of biomechanics, 2016. **49**(9): p. 1562-1569.
34. Hill, M.R., et al., *A theoretical and non-destructive experimental approach for direct inclusion of measured collagen orientation and recruitment into mechanical models of the artery wall.* Journal of biomechanics, 2012. **45**(5): p. 762-771.
35. Weisbecker, H., M.J. Unterberger, and G.A. Holzapfel, *Constitutive modelling of arteries considering fibre recruitment and three-dimensional fibre distribution.* Journal of The Royal Society Interface, 2015. **12**(105): p. 20150111.

36. Dalbosco, M., et al., *Multiscale numerical analyses of arterial tissue with embedded elements in the finite strain regime.* Computer Methods in Applied Mechanics and Engineering, 2021. **381**: p. 113844.
37. Saeidi, S., et al., *Histology-informed multiscale modeling of human brain white matter.* Scientific Reports, 2023. **13**(1): p. 19641.
38. Kapeliotis, M., et al., *Collagen fibre orientation in human bridging veins.* Biomechanics and Modeling in Mechanobiology, 2020. **19**(6): p. 2455-2489.
39. Canham, P., et al., *Medial collagen organization in human arteries of the heart and brain by polarized light microscopy.* Connective tissue research, 1991. **26**(1-2): p. 121-134.
40. Robertson, A.M., et al., *Diversity in the strength and structure of unruptured cerebral aneurysms.* Annals of biomedical engineering, 2015. **43**: p. 1502-1515.
41. Thunes, J.R., et al., *Structural modeling reveals microstructure-strength relationship for human ascending thoracic aorta.* Journal of biomechanics, 2018. **71**: p. 84-93.
42. Stylianopoulos, T. and V.H. Barocas, *Multiscale, structure-based modeling for the elastic mechanical behavior of arterial walls.* 2007.
43. Cavinato, C., et al., *Does the knowledge of the local thickness of human ascending thoracic aneurysm walls improve their mechanical analysis?* Frontiers in bioengineering and biotechnology, 2019. **7**: p. 169.

## Supplementary Materials

To evaluate the robustness of our findings, three additional parametric studies were performed, as summarized in **Table S1**. For each of the studies, Table S1 provides an explanation of the parameters that are varied and identifies the figures containing relevant results. The collective parametric results for stretch differential are shown as bar plots in **Fig. 5**.

**Table S1:** Summary of Parametric Studies Included in Supplementary Materials

| Parametric Study | Varied parameter | Figures |
|---|---|---|
| S1: Thickness Gradient Steepness | Gradual (a = $2.5\times10^4$) | Supp. Fig. S1.1 |
| | Baseline (a = $5\times10^4$) | Fig. 4 |
| | Steep (a = $1\times10^5$) | Supp. Fig. S1.2 |
| S2: Fiber family Orientation (Angle between both fiber families) | 0° (Single fiber family) | Fig. 4 |
| | 30° (± 15° from loading axis) | Supp. Fig. S2.1 |
| | 45° (± 22.5° from loading axis) | Supp. Fig. S2.2 |
| | 60° (± 30° from loading axis) | Supp. Fig. S2.3 |
| S3: Mixed fiber architecture (% of Continuous fibers) | 0% (fully abrupt) | Fig. 4 (a, c) |
| | 25% | Supp. Fig. S3.1 (a, d) |
| | 50% | Supp. Fig. S3.1 (b, e) |
| | 75% | Supp. Fig. S3.1 (c, f) |
| | 100% (fully continuous) | Fig. 4 (b, d) |

### S1 Influence of Thickness Gradient Steepness on Stretch and Stress Fields

To evaluate the sensitivity of our findings to the geometric profile of the wall thickness transition, we simulated two additional gradient levels: a gradual transition (a = $2.5\times10^4$) and a steep transition (a = $1\times10^5$).

**Case S1.A: Gradual Thickness Gradient**

Finite element simulations of the MSM with a gradual thickness gradient under uniaxial tensile loading revealed similar stretch and stress patterns (Supp. Fig. S1.1). At an applied tissue stretch of $\lambda$=1.5, the abrupt fiber termination network exhibited substantial stretch and stress disparities between thick and thin regions. The thin region experienced an average stretch of 1.58 ± 0.09, accompanied by a relatively high von Mises stress of 0.58 ± 0.16 MPa. In contrast, the thick region experienced a lower stretch (1.3 ± 0.17) and stress (0.31 ± 0.14 MPa). The maximum von Mises stress across the entire MSM tissue with the abrupt fiber network was 2.33 MPa. In contrast, the continuously smooth fiber transition network diminished the effect of spatial gradient in the mechanical response. The thin region exhibited an average stretch of 1.48 ± 0.007 and average stress of 0.455 ± 0.11 MPa, while the thick region showed an average stretch of 1.47 ± 0.027 and average stress of 0.453 ± 0.288 MPa. The maximum von Mises stress in the MSM tissue with the continuous fiber network was 0.64 MPa. The percent difference in stretch between both thickness regions decreased from 21.51% for the abrupt fiber termination network to 0.53% for the continuous fiber network.

Fiber-level stresses at $\lambda$=1.5 showed consistent trends. In the abrupt network, collagen fibers in the thin region experienced substantially higher stress (14.64 ± 2.14 MPa) compared to those in the thick region (8.78 ± 6.00 MPa). The continuous network exhibited a more uniform fiber stress distribution, with the thin and thick regions experiencing similar values of 13.95 ± 0.57 MPa and 13.77 ± 0.97 MPa, respectively.

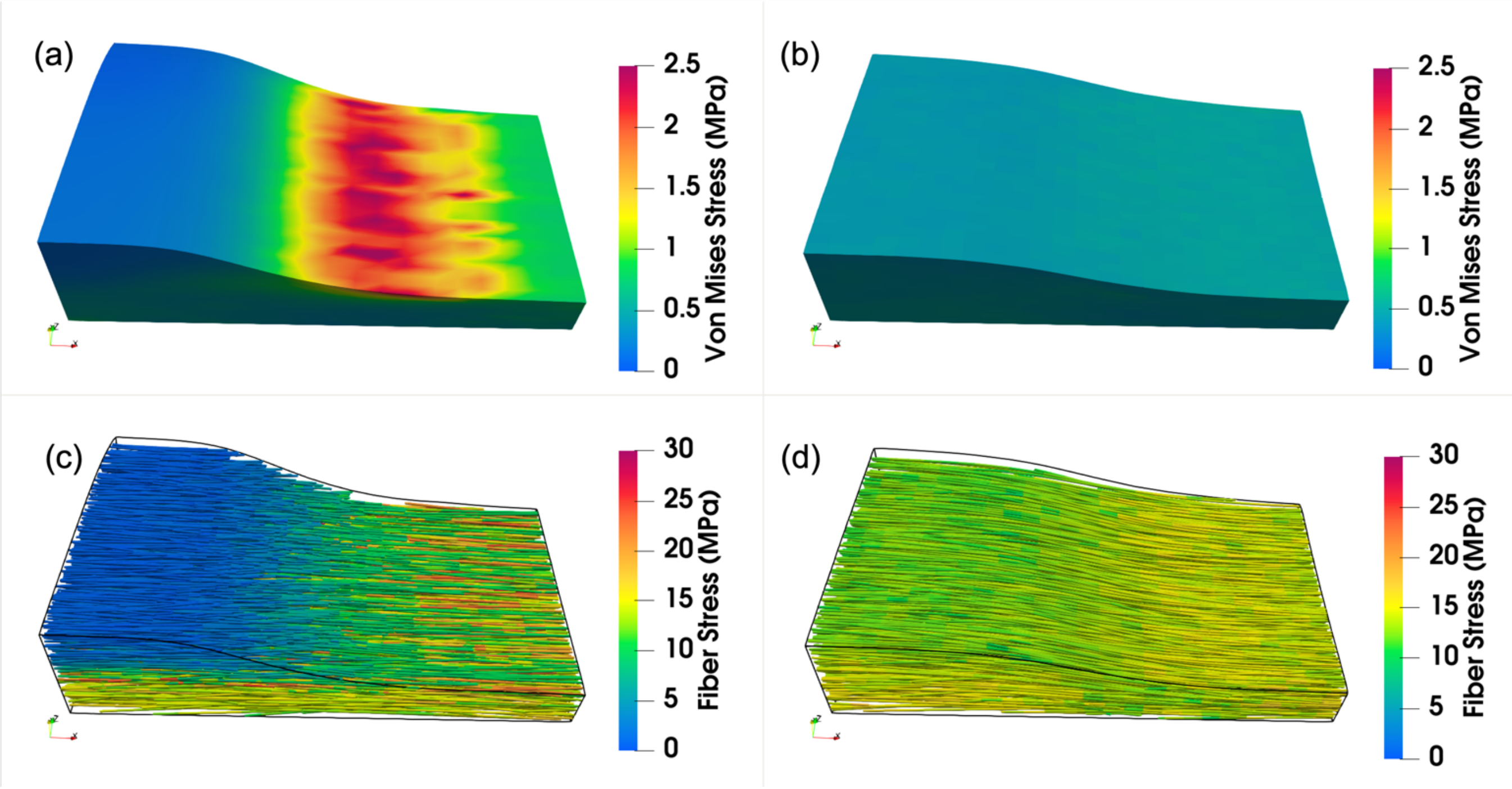

**Supplemental Figure S1.1**: Stress distributions in the meso-scale model with a gradual thickness gradient (a = $2.5\times10^4$) at an applied tissue stretch of $\lambda$ = 1.5. (a, b) Matrix von Mises stress distribution for the (a) Abrupt fiber termination network and (b) Continuous fiber transition network. (c, d) Fiber stress distribution for the (c) Abrupt and (d) Continuous networks. Even with a smoother geometric transition, the abrupt fiber network exhibits stress concentrations in the thin region, whereas the continuous fiber network maintains a highly uniform stress field in both the matrix and fibers.

**Case S1.B: Steep Thickness Gradient**

Finite element simulations of the MSM with a steep thickness gradient under uniaxial tensile loading revealed distinct stretch and stress patterns between the abrupt and smooth fiber networks (Supp. Fig. S1.2). At an applied tissue stretch of $\lambda$=1.5, the abrupt fiber termination network exhibited substantial stretch and stress disparities between thick and thin regions. The thin region experienced an average stretch of 1.56 ± 0.05, accompanied by a relatively high von Mises stress of 0.54 ± 0.058 MPa. In contrast, the thick region experienced a lower stretch (1.29 ± 0.19) and stress (0.33 ± 0.14 MPa). The maximum von Mises stress across the entire MSM tissue with the abrupt fiber network was 1.86 MPa. In contrast, the continuously smooth fiber transition network diminished the spatial gradient in the mechanical response. The thin region exhibited an average stretch of 1.46 ± 0.013 and average stress of 0.43 ± 0.015 MPa, while the thick region showed an average stretch of 1.45 ± 0.057 and average stress of 0.42 ± 0.057 MPa. The maximum von Mises stress in the MSM tissue with the continuous fiber network was 1.15 MPa. The percent difference in stretch between the two constant thickness regions decreased from 21.02% for the abrupt fiber termination network to 0.79% for the continuous fiber network. Fiber-level stresses at $\lambda$=1.5 followed a similar pattern. In the abrupt network, collagen fibers in the thin region experienced substantially higher stress (14.65 ± 2.20 MPa) compared to those in the thick region (7.99 ± 6.17 MPa). The continuous network exhibited a more uniform fiber stress distribution, with the thin and thick regions experiencing similar values of 13.11 ± 0.65 MPa and 12.67 ± 1.81 MPa, respectively.

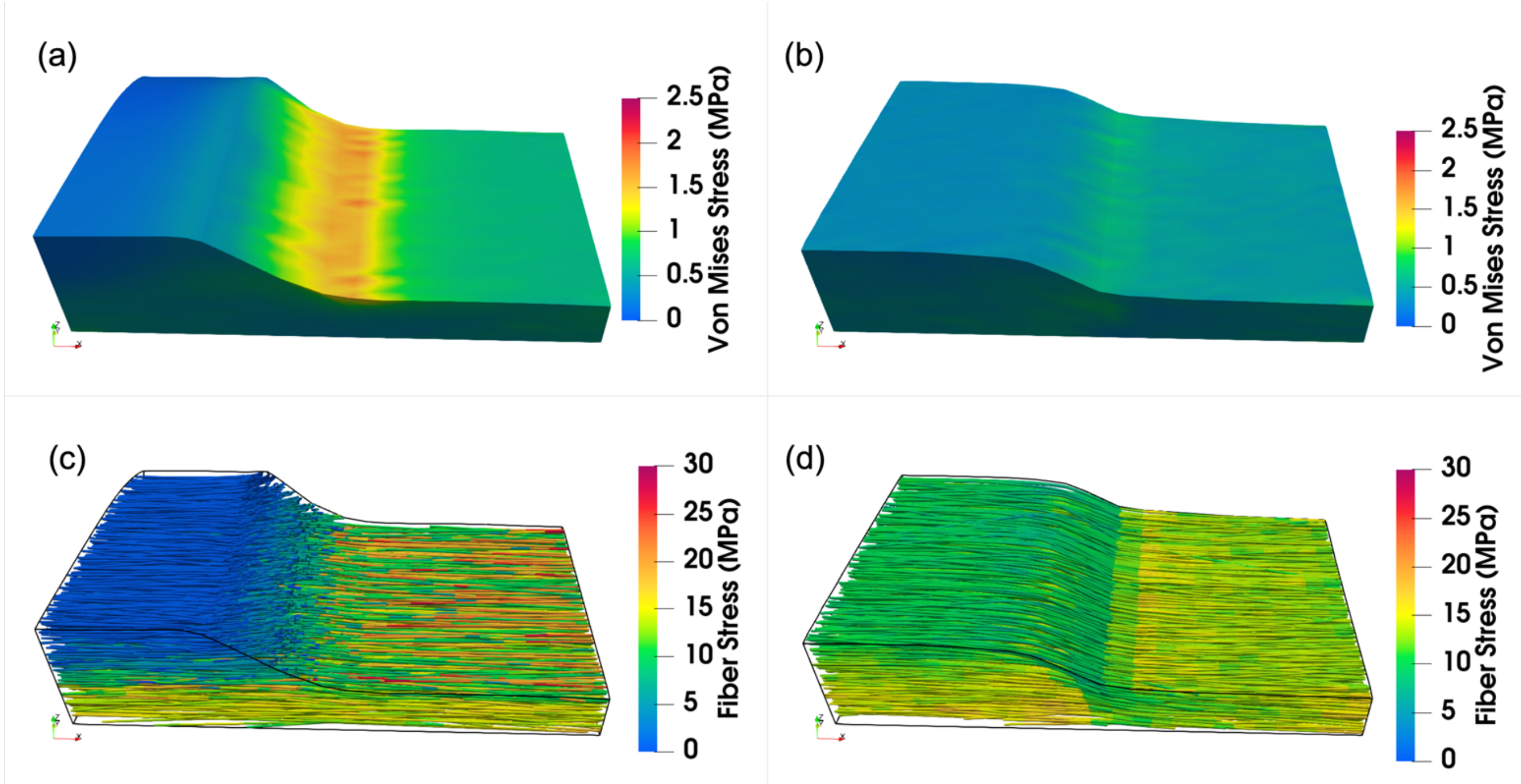

**Supplemental Figure S1.2**: Stress distributions in the meso-scale model with a steep thickness gradient (a = 10×10$^4$) at an applied tissue stretch of $\lambda$ = 1.5. (a, b) Matrix von Mises stress distribution for the (a) Abrupt fiber termination network and (b) Continuous fiber transition network. (c, d) Fiber stress distribution for the (c) Abrupt and (d) Continuous networks. Even for steep gradient in the thickness transition region, the continuous fiber network effectively mitigates these geometric effects, resulting in a homogenized stress distribution.

## S2 Influence of Fiber Orientation (Dual Fiber Families)

To investigate the influence of fiber orientation on the mechanical response, we modeled tissues with dual fiber families symmetric about the circumferential loading direction. We examined three distinct architectures with total included angles between fiber families of 30° (± 15° from loading axis), 45° (± 22.5°), and 60° (± 30°).

**Case S2.A: 30° Total Angle (± 15° Orientation from loading direction)**

For the highly aligned network (30° total angle), fibers provided significant resistance to the axial load (Supp. Fig. S2.1). In the abrupt fiber termination network, stretch and stress disparities were observed between the regions. The thin region experienced an average stretch of 1.57 ± 0.06, accompanied by a von Mises stress of 0.56 ± 0.07 MPa. In contrast, the thick region experienced a lower stretch (1.30 ± 0.17) and stress (0.34 ± 0.13 MPa). In contrast, the continuous fiber transition network effectively homogenized the mechanical field. The thin region exhibited an average stretch of 1.48 ± 0.009 and average stress of 0.454 ± 0.010 MPa, while the thick region showed an average stretch of 1.46 ± 0.04 and average stress of 0.446 ± 0.037 MPa. The stretch difference between thick and thin regions decreased from 20.46% (abrupt) to 0.8% (continuous).

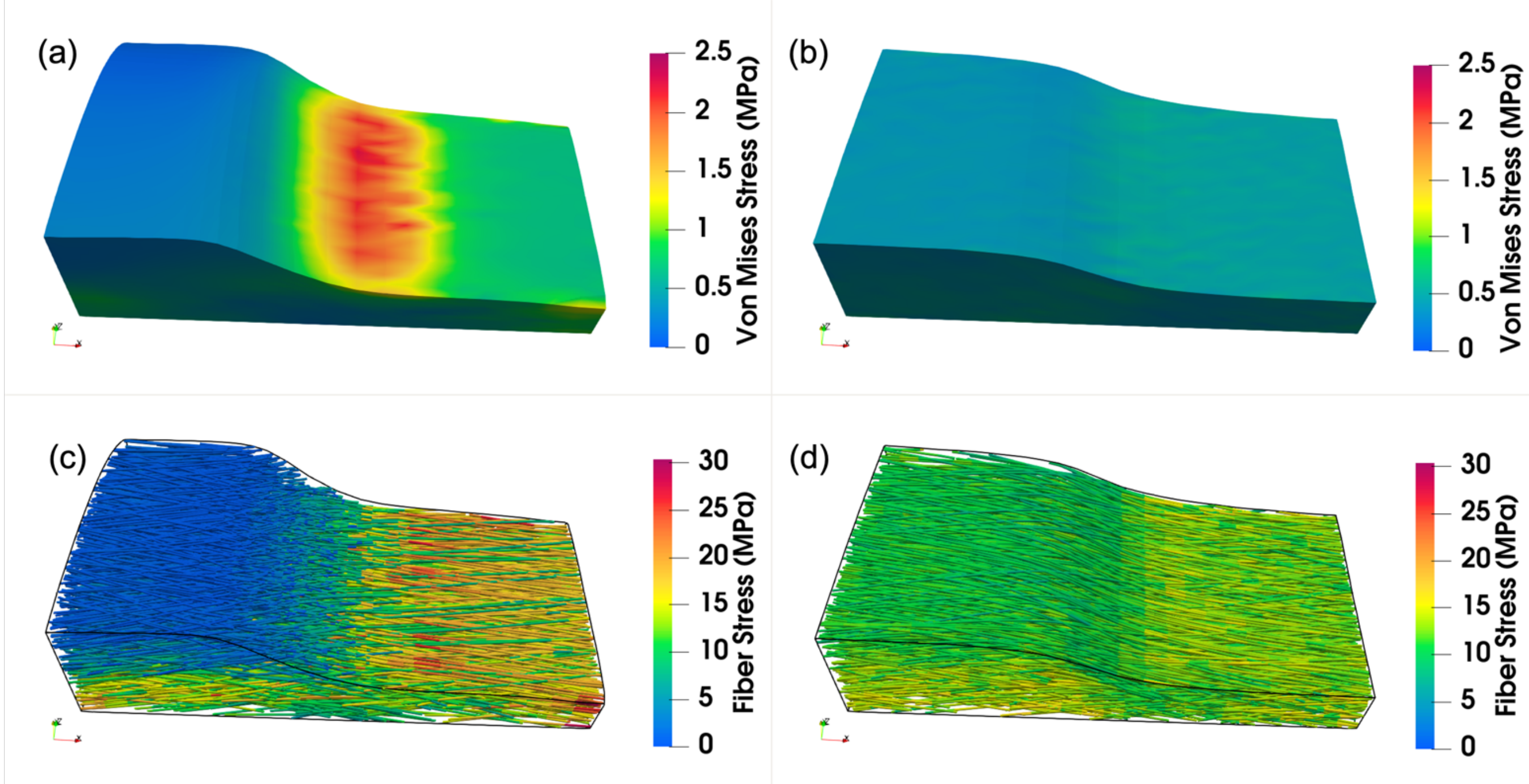

**Supplementary Figure S2.1**: Mechanical response for dual fiber families with a total included angle of 30° (±15° relative to loading) at an applied tissue stretch of $\lambda$=1.5. (a, b) Spatial distribution of tissue stretch for the (a) Abrupt fiber termination and (b) Continuous fiber transition networks. (c, d) Fiber stress distribution for the (c) Abrupt and (d) Continuous networks. High fiber alignment allows the continuous network to effectively bridge the transition, preventing the stretch disparities and fiber stress concentrations observed in the thin region of the abrupt network.

**Case S2.B: 45˚ Total Angle (± 22.5˚ Orientation from loading direction)**

As the total angle increased to 45˚, the fibers became slightly less aligned with the loading axis (Supp. Fig. S2.2). In the abrupt fiber termination network, the thin region experienced an average stretch of 1.57 ± 0.06 and a von Mises stress of 0.57 ± 0.071 MPa. The thick region showed a stretch of 1.32 ± 0.17 and stress of 0.37 ± 0.12 MPa, resulting in a stretch differential of 19.2%. The continuous fiber transition network maintained a uniform response despite the change in angle. The thin region exhibited an average stretch of 1.48 ± 0.01 and stress of 0.45 ± 0.011 MPa, while the thick region showed an average stretch of 1.46 ± 0.04 and stress of 0.45 ± 0.038 MPa. The stretch differential remained low at 0.86%.

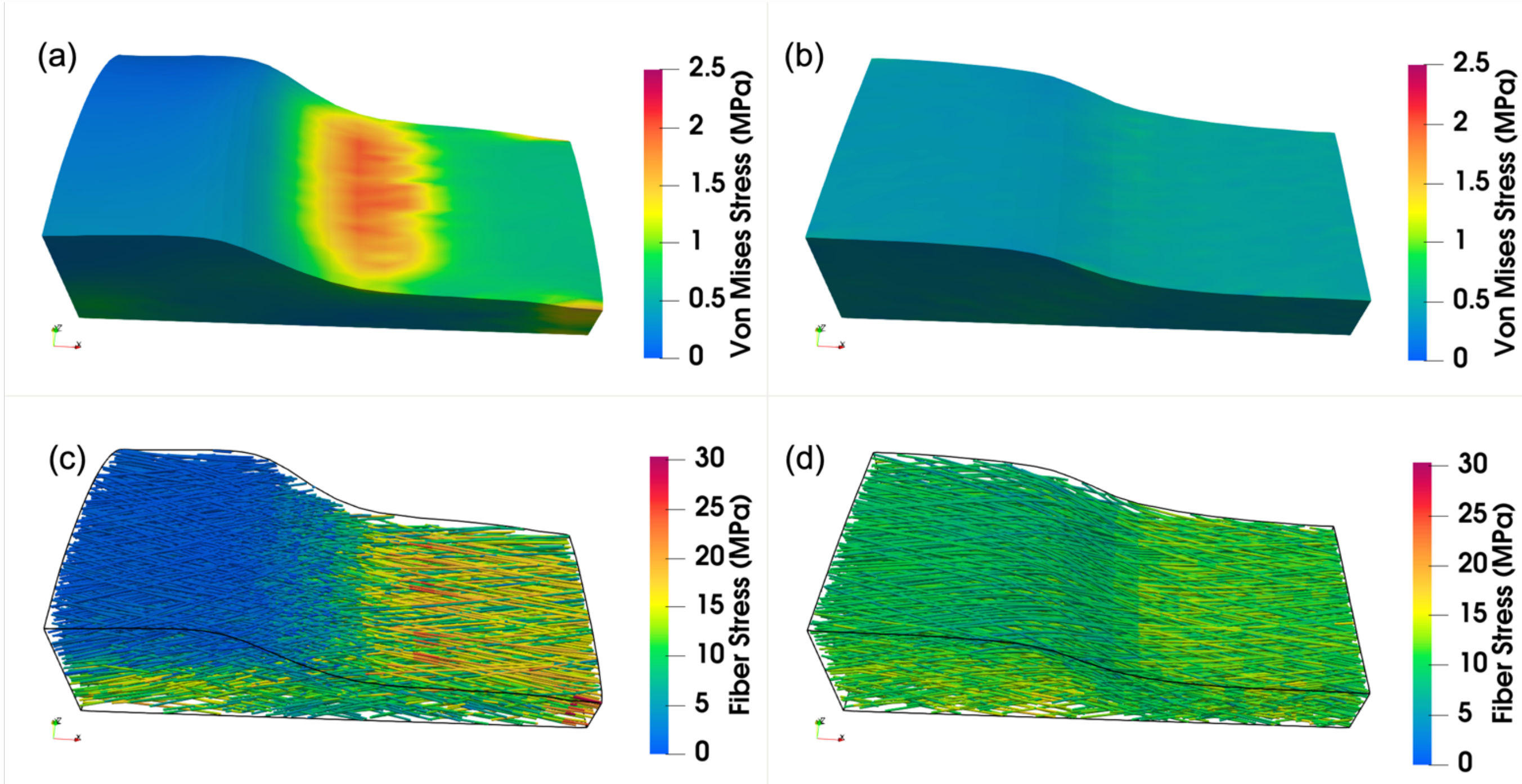

**Supplementary Figure S2.2**: Mechanical response for dual fiber families with a total included angle of 45° (±22.5° relative to loading) at an applied tissue stretch of $\lambda$=1.5. (a, b) Spatial distribution of tissue stretch for the (a) Abrupt fiber termination and (b) Continuous fiber transition networks. (c, d) Fiber stress distribution for the (c) Abrupt and (d) Continuous networks. Despite reduced axial alignment, the continuous network maintains a uniform strain field and fiber stress distribution, whereas the abrupt network exhibits distinct concentrations in the thinner region.

**Case S2.C: 60° Total Angle (± 30° Orientation from loading direction)**

At a total angle of 60°, the fibers had the largest deviation from the loading axis in this study (Supp. Fig. S2.3). In the abrupt fiber termination network, the mechanical disparity was most pronounced. The thin region experienced an average stretch of 1.58 ± 0.06 and a relatively high stress of 0.576 ± 0.072 MPa, compared to the thick region (stretch: 1.34 ± 0.15; stress: 0.38 ± 0.11 MPa). The stretch disparity between regions reached 18.15%. However, even at this higher orientation angle, the continuous fiber transition network acted as a "safety net," redistributing the load. The thin region exhibited an average stretch of 1.48 ± 0.01 and stress of 0.45 ± 0.009 MPa, while the thick region showed an average stretch of 1.46 ± 0.04 and stress of 0.447 ± 0.037 MPa. The stretch differential was reduced to 0.98%, confirming that fiber continuity mitigates geometric stress concentrations across a range of physiological fiber angles.

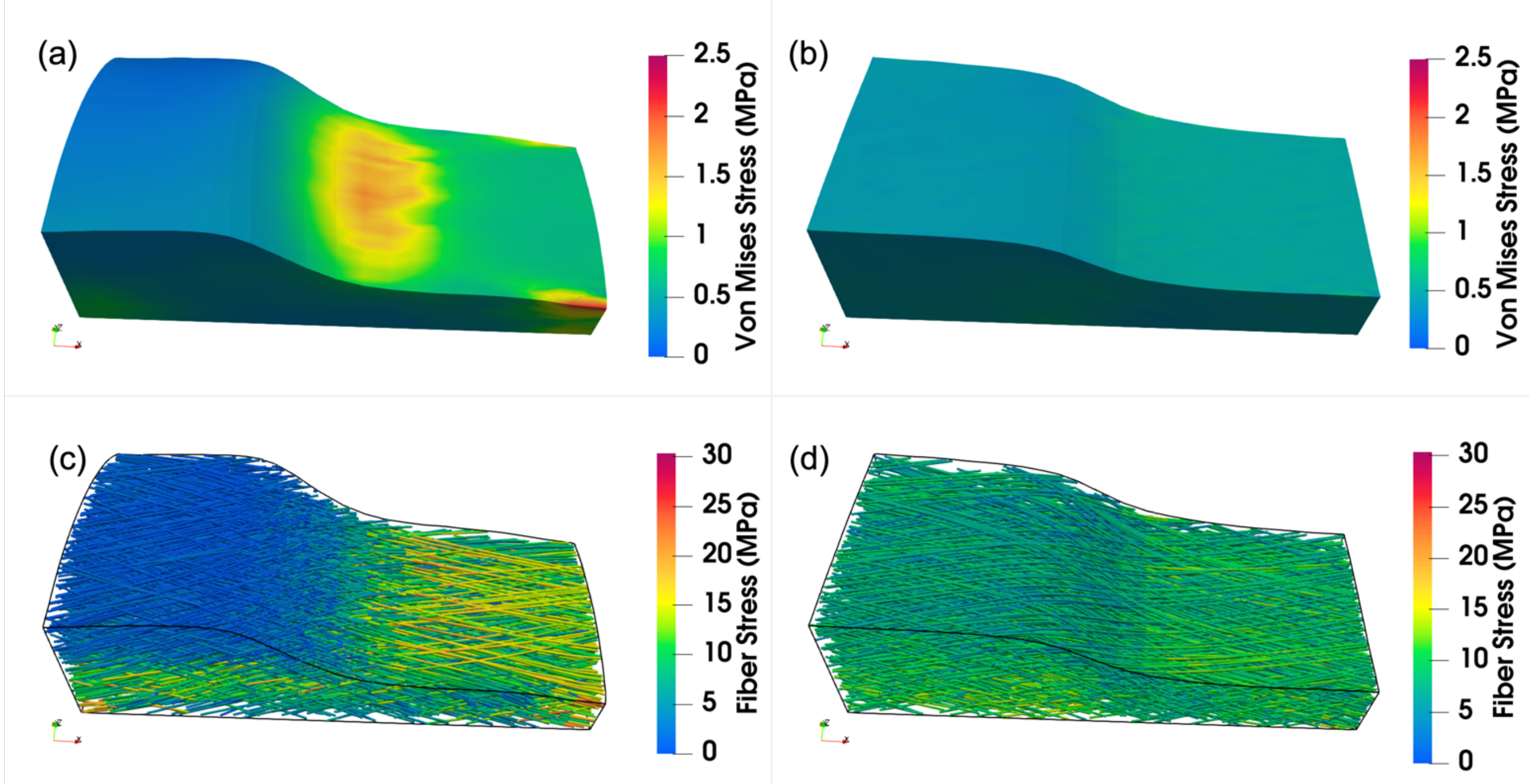

**Supplementary Figure S2.3**: Mechanical response for dual fiber families with a total included angle of $60° (±30° relative to loading) at an applied tissue stretch of $\lambda$=1.5. (a, b) Spatial distribution of tissue stretch for the (a) Abrupt fiber termination and (b) Continuous fiber transition networks. (c, d) Fiber stress distribution for the (c) Abrupt and (d) Continuous networks. Even with significant fiber deviation from the loading axis, fiber continuity effectively mitigates the pronounced geometric stretch disparities and fiber stress concentrations characteristic of the abrupt termination network.

## S3 Influence of Mixed Fiber Architectures

In native tissues, it is possible that fiber architecture is not binary (i.e., purely abrupt or purely continuous) but rather exists as a hybrid mixture where a fraction of fibers bridges the thickness gradient while others terminate. To investigate this physiological intermediate state, we simulated networks with varying ratios of continuous-to-abrupt fibers: 0% (purely abrupt), 25%, 50%, 75%, and 100% (purely continuous) as shown in Supp. Fig. S3.1. The FEM element results of 0% and 100% are already shown in our primary analysis (Fig. 4).

**Case S3.A: 25% Continuous Fibers**

In this network, only one-quarter of the fibers were continuous, while the majority (75%) terminated abruptly. The mechanical response remained dominated by the geometric effect of wall thinning. The thin region experienced an average stretch of 1.56 ± 0.05 and a von Mises stress of 0.54 ± 0.066 MPa. The thick region showed a lower stretch (1.34 ± 0.14) and stress (0.34 ± 0.12 MPa). However, even this small fraction of continuous fibers reduced the stretch differential to 16.4% (down from ~21% in the purely abrupt case).

**Case S3.B: 50% Continuous Fibers**

With an equal mix of continuous and abrupt fibers, the "bridging" effect became more pronounced. The stretch in the thin region decreased to 1.54 ± 0.04 and a von Mises stress of 0.51 ± 0.049 MPa, while the stretch in thick region increased slightly to 1.38 ± 0.11 with von Mises stress of 0.38 ± 0.098 MPa. The stretch differential dropped to 10.97%, effectively halving the disparity observed in the abrupt model. This confirms that the load-transfer benefit scales almost linearly with the proportion of continuous fibers.

**Case S3.C: 75% Continuous Fibers**

When the majority of fibers (75%) were continuous, the mechanical field approached homogeneity. The thin region stretch was 1.51 ± 0.02 and the von Mises stress was 0.48 ± 0.029 MPa while thick region experienced the average stretch of 1.43 ± 0.08 with average von Mises stress of 0.41 ± 0.073 MPa. The stretch differential was further reduced to 5.75%, indicating that a fully continuous network is not strictly required to achieve a mechanically robust, nearly uniform response; a sufficiently high-volume fraction of bridging fibers can largely override geometric stress concentrations.

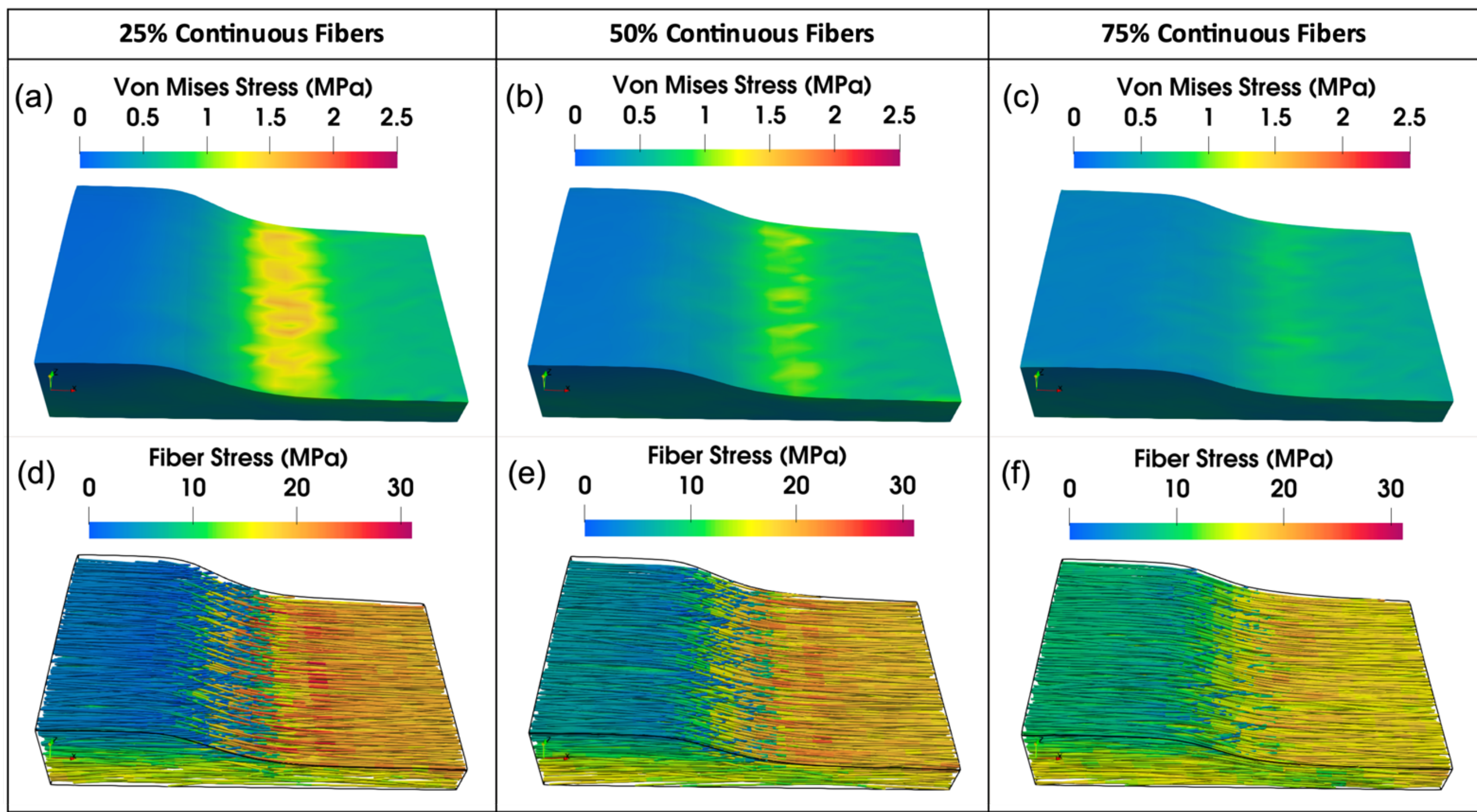

**Supplementary Figure S3.1:** Stress distributions in the meso-scale model for mixed fiber architectures at an applied tissue stretch of $\lambda$=1.5. Columns display results for networks with (a, d) 25% continuous fibers, (b, e) 50% continuous fibers, and (c, f) 75% continuous fibers. The top row (a, b, c) shows the Matrix von Mises stress, while the bottom row (d, e, f) shows the fiber stress. As the percentage of continuous fibers increases from 25% to 75%, the stress concentrations in the thin region progressively diminish, resulting in a more uniform stress distribution across the entire tissue volume.

The results revealed a monotonic, nearly linear relationship between the percentage of continuous fibers and the homogenization of the strain field (Fig. 5c). The stretch differential between thick and thin regions decreased progressively from **20.97%** in the purely abrupt network (0% continuous) to 16.4 at 25%, 10.97% at 50%, and 5.75% at 75%, finally reaching **0.68%** in the purely continuous network. This suggests that even a partial preservation of fiber continuity confers significant mechanical benefits. For instance, having just 50% of fibers span the defect reduced the stretch disparity by approximately half, indicating that the "bridging" effect is additive and proportional to the volume fraction of continuous load paths.

# S4 The Stress-Stress Plots for all Tortuosity Cases

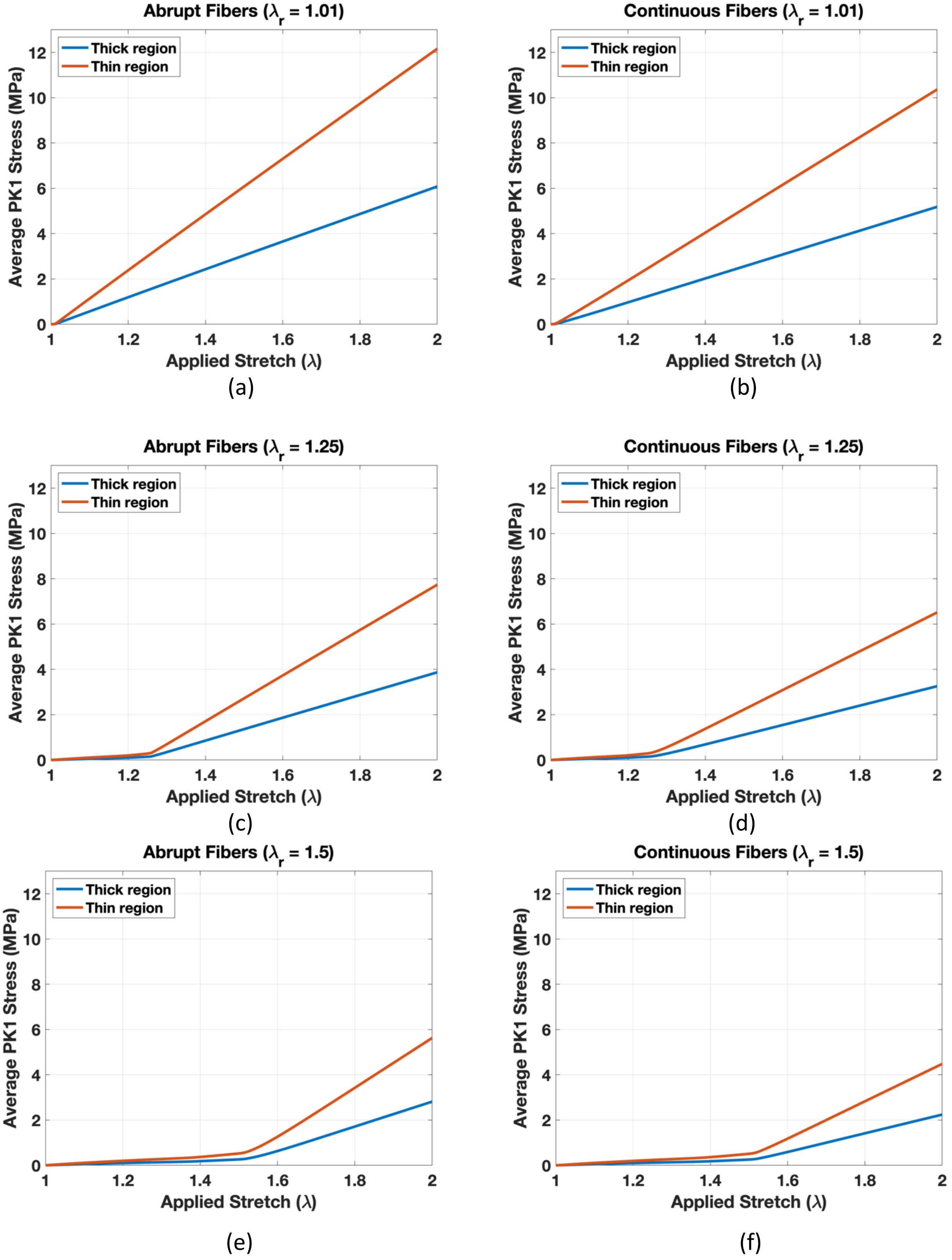

**Supplemental Figure S4.1**: First Piola-Kirchhoff stress–stretch curves for thin and thick regions under different fiber architectures and recruitment stretches. (a, c, e) Abrupt fiber terminations; (b, d, f) Continuous fibers. Top row (a, b) corresponds to low recruitment stretch ($\lambda_r$ = 1.01), middle row (c, d) corresponds to delayed fiber recruitment ($\lambda_r$ = 1.25), and bottom row (e, f) corresponds to delayed fiber recruitment ($\lambda_r$ = 1.5).